\documentclass[sn-basic,linktocpage,pdflatex]{sn-jnl}
\usepackage{hyperref}
\usepackage{graphicx}%
\usepackage{multirow}%
\usepackage{amsmath,amssymb,amsfonts}%
\usepackage{amsthm}%
\usepackage{mathrsfs}%
\usepackage[title]{appendix}%
\usepackage{xcolor}%
\usepackage{textcomp}%
\usepackage{manyfoot}%
\usepackage{booktabs}%
\usepackage{algorithm}%
\usepackage{algorithmicx}%
\usepackage{algpseudocode}%
\usepackage{listings}%
\usepackage{array, makecell}
\usepackage{bm}
\usepackage{xspace}
\usepackage{orcidlink}
\setcitestyle{authoryear,open={(},close={)}}

\graphicspath{figures}

\usepackage{aas_macros}

\theoremstyle{thmstyleone}%
\theoremstyle{thmstyletwo}%

\theoremstyle{thmstylethree}%

\begin{document}

\title[evolution of switchbacks]{Evolution and Impact of Switchbacks Throughout the Heliosphere}


\author[1]{\fnm{Alfred} \sur{Mallet} \orcidlink{0000-0001-9202-1340}}
\equalcont{These authors contributed equally to this work.}

\author[2,3]{\fnm{Chen} \sur{Shi} \orcidlink{0000-0002-2582-7085}} 
\equalcont{These authors contributed equally to this work.}

\author[4]{\fnm{Anna} \sur{Tenerani} \orcidlink{0000-0003-2880-6084}}
\equalcont{These authors contributed equally to this work.}


\author[1, 25]{\fnm{Oleksiy} \sur{Agapitov} \orcidlink{0000-0001-6427-1596}}

\author[5]{\fnm{Mojtaba} \sur{Akhavan-Tafti} \orcidlink{0000-0003-3721-2114}}

\author[6]{\fnm{Samuel} \sur{Badman} \orcidlink{0000-0002-6145-436X}}

\author[7]{\fnm{Nina} \sur{Bizien} \orcidlink{0000-0001-6767-0672}}

\author[1]{\fnm{Trevor} \sur{Bowen} \orcidlink{0000-0002-4625-3332}}

\author[8]{\fnm{Mihir I.} \sur{Desai} \orcidlink{0000-0002-7318-6008}}

\author[9]{\fnm{J. F.} \sur{Drake} \orcidlink{0000-0002-9150-1841}}

\author[10]{\fnm{Timothy} \sur{Horbury} \orcidlink{0000-0002-7572-4690}}

\author[11]{\fnm{Andrea} \sur{Larosa} \orcidlink{0000-0002-7653-9147}}

\author[12,13,14]{\fnm{Maria S.} \sur{Madjarska} 
\orcidlink{0000-0001-9806-2485}}

\author[15]{\fnm{Francesco} \sur{Malara} \orcidlink{0000-0002-5554-8765}}

\author[10]{\fnm{Lorenzo} \sur{Matteini} \orcidlink{0000-0002-6276-7771}}

\author[16]{\fnm{Mathew} \sur{Owens} \orcidlink{0000-0003-2061-2453}}

\author[17]{\fnm{Victor} \sur{R\'eville} \orcidlink{0000-0002-2916-3837}}

\author[1]{\fnm{Nikos} \sur{Sioulas} \orcidlink{0000-0002-1128-9685}}

\author[5]{\fnm{Shirsh Lata} \sur{Soni} \orcidlink{0000-0002-5550-739X}}

\author[20]{\fnm{Jonathan} \sur{Squire} \orcidlink{0000-0001-8479-962X}}

\author[21]{\fnm{Gabriel Ho Hin} \sur{Suen} \orcidlink{0000-0002-9387-5847}}

\author[9]{\fnm{Marc} \sur{Swisdak} \orcidlink{0000-0002-5435-3544}}

\author[3]{\fnm{Marco} \sur{Velli} \orcidlink{0000-0002-2381-3106}}

\author[22]{\fnm{Jaye} \sur{Verniero} \orcidlink{0000-0003-1138-652X}}

\author[23,24]{\fnm{Nicholas} \sur{Watkins}}


\author*[18,11,19]{\fnm{Luca} \sur{Sorriso-Valvo} \orcidlink{0000-0002-5981-7758}}\email{lucsv@kth.se}


\affil[1]{\orgdiv{Space Sciences Laboratory}, \orgname{University of California, Berkeley}, \orgaddress{ \city{Berkeley}, \state{California}, \country{USA}}}

\affil[2]{\orgdiv{Department of Physics}, \orgname{Auburn University}, \orgaddress{\city{Auburn}, \state{Alabama}, \postcode{36849}, \country{USA}}}

\affil[3]{\orgdiv{Department of Earth, Planetary, and Space Sciences}, \orgname{University of California, Los Angeles}, \orgaddress{\city{Los Angeles}, \state{California}, \postcode{90095}, \country{USA}}}

\affil[4]{\orgdiv{Department of Physics}, \orgname{The University of Texas at Austin}, \orgaddress{\city{Austin}, \state{Texas}, \country{USA}}}

\affil[5]{\orgdiv{Department of Climate and Space Sciences and Engineering}, \orgname{University of Michigan}, \orgaddress{\street{2455 Hayward St.}, \city{Ann Arbor}, \state{Michigan}, \postcode{48109}, \country{USA}}}

\affil[6]{\orgdiv{Center for Astrophysics}, \orgname{Harvard \& Smithsonian}, \orgaddress{\city{Cambridge}, \state{Massachusetts}, \postcode{02138}, \country{USA}}}

\affil[7]{\orgdiv{LPC2E}, \orgname{CNRS/University of Orl\'eans/CNES}, \orgaddress{\city{Orl\'eans}, \country{France} }}

\affil[8]{\orgname{Southwest Research Institute}, \orgaddress{\street{6220 Culebra Road}, \city{San Antonio}, \state{Texas}, \postcode{78238}, \country{USA} }}

\affil[9]{\orgname{University of Maryland College Park}, \orgaddress{\city{College Park}, \state{Maryland}, \country{USA}}}

\affil[10]{\orgdiv{Department of Physics}, \orgname{Imperial College London}, \orgaddress{\city{London}, \country{UK}}}

\affil[11]{\orgdiv{Institute for Plasma Science and Technology (ISTP)}, \orgname{CNR}, \orgaddress{\city{Bari}, \country{Italy}}}


%
\affil[12]{Max Planck Institute for Solar System Research,
Justus-von-Liebig-Weg 3, 37077, G\"ottingen, Germany}
\affil[13]{Korea Astronomy and Space Science Institute, 34055, Daejeon,
Republic of Korea}
\affil[14]{Space Research and Technology Institute, Bulgarian Academy
of Sciences, Acad. G. Bonchev Str., Bl. 1, 1113, Sofia, Bulgaria}

\affil[15]{\orgdiv{Department of Physics}, \orgname{University of Calabria}, \orgaddress{\country{Italy}}}

\affil[16]{\orgdiv{Department of Meteorology}, \orgname{University of Reading}, \orgaddress{\city{Reading}, \country{UK}}}

\affil[17]{\orgdiv{IRAP}, \orgname{Université Toulouse III—Paul Sabatier, CNRS, CNES} \orgaddress{\city{Toulouse}, \country France}}


\affil*[18]{\orgdiv{Department of Electromagnetics and Plasma Physics, School of Electrical Engineering and Computer Science}, \orgname{KTH - Royal Institute of Technology}, \orgaddress{\city{Stockholm}, \country{Sweden}}}

\affil[19]{\orgname{Swedish Institute of Space Physics}, \orgaddress{\city{Uppsala}, \country{Sweden}}}

\affil[20]{\orgdiv{Physics Department}, \orgname{University of Otago}, \orgaddress{\city{Dunedin}, \country{New Zealand}}}

\affil[21]{\orgdiv{Department of Space and Climate Physics}, \orgname{University College London}, \orgaddress{\city{London}, \country{UK}}}

\affil[22]{\orgdiv{Heliophysics Science Division}, \orgname{NASA Goddard Space Flight Center}, \orgaddress{\city{Greenbelt}, \state{Maryland}, \postcode{20771}, \country{USA}}}

\affil[23]{\orgdiv{Centre for Fusion Space and Astrophysics}, \orgname{University of Warwick}, \orgaddress{\city{Coventry}, \country{UK}}}

\affil[24]{\orgdiv{Grantham Research Institute on Climate Change and the Environment}, \orgname{London School of Economics and Political Science}, \orgaddress{\city{London}, \country{UK}}}

\affil[25]{\orgname{Astronomy and Space Physics Department, National Taras Shevchenko University of Kyiv}, \orgaddress{\city{Kyiv}, \postcode{01601}, \country{Ukraine}}}








\abstract{Magnetic switchbacks are large-amplitude fluctuations in the interplanetary magnetic field, and appear frequently in the near-Sun solar wind explored recently by Parker Solar Probe: these new observations have prompted many new studies into their properties and origins. Here, we first review what is known about how switchbacks evolve as they travel away from the Sun: both in terms of their expansion-driven growth and their decay due to various processes like turbulence, reconnection, dispersion, parametric instability, and interaction with interplanetary shocks. We then review the current state of knowledge on how switchbacks impact the physics of the solar wind as a whole: in terms of the turbulent cascade, acceleration and heating of the wind, modification of the open solar flux and scattering of energetic particles. Finally, we suggest future studies to further our understanding of switchback evolution and impacts on the heliosphere.}




\maketitle

\tableofcontents

\section{Introduction}\label{sec1}


One of the most surprising observations made by Parker Solar Probe (PSP) \citep{Fox2016,Raouafi2023b}, even in data from its first perihelion \citep{Bale2019,Kasper2019}, is that in the near-Sun solar wind, the magnetic field is not smooth: instead, it is dominated by large-amplitude, impulsive rotations of the magnetic field, dubbed “switchbacks”. 

The discovery of switchbacks prompted a flurry of theoretical, numerical, and observational activity to understand these new observations: such rapid scientific progress had the side-effect that different groups adopted different definitions of switchbacks, and it is debatable as to whether a consensus definition has been reached. Nevertheless, we note briefly here that typically, the switchbacks observed by PSP have (i) large amplitude $\delta B/ B\sim 1$, (ii) small fluctuations in the magnetic field strength $\delta |\boldsymbol{B}|/|\delta \boldsymbol{B}|\ll 1$, and (iii) quasi-Alfv\'enic velocity fluctuations correlated with the magnetic field fluctuations, $\delta \boldsymbol{u} \approx \delta \boldsymbol{B}/\sqrt{4\pi\rho}$, where $\rho$ is the mass density. However, we would like to emphasise that the boundary between switchbacks and the rest of the fluctuations in the solar wind is certainly fuzzy, if it has any meaning at all. 
The switchbacks are preferentially found in Alfv\'enic solar wind streams with high cross-helicity, and may be considered part of the flux of turbulent fluctuations continually emitted by the Sun. A comprehensive account of the properties of switchbacks can be found in \cite{paper3}. 
Whether switchbacks are generated by dynamical processes in the solar wind, or are impulsively generated in the corona, remains a matter of debate \citep{paper4,paper5}.


Irrespective of their precise origin, the existence (and indeed dominance) of the switchbacks in the near-Sun Alfv\'enic solar wind is a clear fact based on the data. The fact that the properties of switchbacks observed further out in the heliosphere (at 1~au and beyond) are very different from the new PSP observations implies that these structures dynamically evolve and decay in the expanding solar wind. It is therefore crucial to understand what impacts switchbacks have on the physics of the solar corona and wind, and on the interpretation of other observations. Before examining these impacts, one must first understand how switchbacks evolve after they are formed: both how they might grow in amplitude relative to the background field and change their morphologies due to the expansion of the solar wind, and also how they might decay due to a wide variety of erosion processes.

In the first part of this review, we will discuss the important effects of solar wind expansion on switchbacks. As switchbacks propagate and are advected through this inhomogeneous medium, their normalized amplitude increases in a characteristic way (Sec.~\ref{sec:amplitudeevo}). 
Moreover, the background magnetic field follows the Parker spiral rather than pointing radially away from the Sun: this introduces important asymmetries into the morphology of the switchbacks, independently of their origin (Sec.~\ref{sec:parkerspiral}).
Once switchbacks attain a large amplitude, this evolution also leads naturally to steepening at their boundaries (Sec.~\ref{sec:boundaries}). Finally, the inhomogeneity of the solar wind also introduces wave reflections, producing a population of inward propagating Alfv\'enic fluctuations leading to a characteristic evolution of Alfv\'enicity with radial distance, and a turbulent cascade of energy (Sec.~\ref{sec:alfvenicity}). 

While the solar wind expansion may cause switchbacks to reach large 
amplitude as compared to the background magnetic field, it is likely that this growth does not continue inevitably: at some point, various non-ideal and/or dissipative processes may occur, regulating the switchback amplitude and morphology. Which processes are important to this regulation is still a matter of open research, and we discuss the existing literature in Sec.~\ref{sec:nonideal}. First, the fact that the switchbacks exhibit such steep and discontinuous boundaries means that various kinetic effects may be important. For example, new nonlinearities and the dispersive nature of plasma waves at scales comparable to the ion kinetic scales means that switchbacks may not be stable once their boundaries become sufficiently steep, emitting dispersive waves as part of their decay (although in some situations the boundaries may stabilize into steady nonlinear solitary waves) (Sec.~\ref{sec:nonideal_boundaries}). Moreover, the important process of magnetic reconnection at the boundaries may efficiently convert the magnetic energy of the switchbacks into heat (Sec.~\ref{sec:reconnection}). Another possibility is the parametric decay of switchbacks. This is naturally expected to be important due to the large-amplitude waves, but may be suppressed due to the localized nature of the switchbacks (Sec.~\ref{sec:pdi}). This process produces a backward-propagating Alfv\'en wave as well as compressive fluctuations, and therefore may lead to the development of a robust turbulent cascade. Finally, it is also of interest to study the processing of switchbacks by interplanetary shocks: they are deformed due to the interaction which may have important implications for their subsequent evolution and the dynamics of the plasma heating around the shock (Sec.~\ref{sec:IP_shocks}).

Given this information on how switchbacks evolve and decay, what can we say about their impact on the physics of the solar wind and corona as a whole? We showcase some early work on this subject in Sec.~\ref{sec:impacts}. First, one generic way in which switchbacks can affect the solar wind is through the development of turbulence and associated heating and acceleration of the plasma: indeed, it is still a matter of debate whether switchbacks comprise a separate and distinct population from the rest of the turbulence. In any case, it is evident that switchbacks (depending, of course, on their precise definition -- see \citet{paper3}) constitute a large part of the total fluctuation power, and thus possibly provide an important source of free energy for this process -- see Secs.~\ref{sec:turbulence}, \ref{sec:acceleration}, and \ref{sec:heating}.
Finally, due to their remarkably discontinuous boundaries and large amplitudes, switchbacks can scatter particles very efficiently (Sec.~\ref{sec:particles}), meaning that the statistics of the switchback population must be taken into account for realistic modeling of energetic particle transport in the solar system.

\section{Effects of Solar Wind Expansion on Switchbacks}
\label{sec:expansion}

\subsection{Theory and Observations of the Radial Evolution of Wave Energy}\label{sec:amplitudeevo}

Essential to understanding switchbacks is the theory of wave propagation in an inhomogeneous flow. In this case, the wave energy is not conserved, as the wave pressure does work on the flow, and the conservation of the wave energy is generalized to the conservation law of the wave action $\mathcal S=\mathcal E/\omega$, where $\mathcal{E}$ is the wave energy density and $\omega$ its intrinsic frequency. The theory of wave action conservation has been known for a long time, starting from the work of \citet{witham1965general} and \citet{bretherton1968wavetrains} based on the so-called Wentzel-Kramers-Brillouin (WKB) approximation. The WKB approximation assumes that the length-scale over which the underlying medium is changing is much larger than the wavelength of the wave. This limit allows one to neglect effects such as wave reflection due to inhomogeneities, leading to the following conservation law:
\begin{equation}
\frac{\partial \mathcal S}{\partial t}+\nabla\cdot({\bf v_g}\mathcal S)=0,
\label{eq:wave_action}
\end{equation}
where ${\bf v_g}$ is the group velocity of the wave. The conservation law of wave action is of general validity, but here we will discuss its implications for Alfvén waves in the solar wind, assuming a radial mean magnetic field and flow. We nevertheless mention that the conservation law expressed in Eq. (\ref{eq:wave_action}) can be generalized to include finite wavelength effects to go beyond the WKB approximation \citep{heinemann1980non,velli2000alfvenic,chandran2015conservation}. Then, in the case of Alfvén waves, the total wave action of forward and backward propagating waves must be conserved. Recently, a generalization of total wave action conservation to include couplings with magnetosonic modes has also been discussed, relevant when $\beta\simeq 1$  and mode degeneracy occurs \citep{huang2022conservation}. The effects of non-radial magnetic fields are discussed in Sec.~\ref{sec:parkerspiral}. 

Let us consider now a steady-state, radially expanding solar wind with velocity ${\bf U}=U(R)\hat{\boldsymbol{e}}_R$, Alfv\'en speed ${\bf V_a}=B_r(R)/\sqrt{4\pi\rho(R)}\hat{\boldsymbol{e}}_R$, and mass density $\rho(R)$, where $R$ is the radial distance from the Sun and $\hat{\boldsymbol{e}}_R$ is the unit vector in the radial direction. Then, for Alfvén waves with velocity perturbation ${\bf u}$, Eq.~(\ref{eq:wave_action}) yields
\begin{equation}
u^2 \frac{(U+V_a)^2}{UV_a}=const.
\label{eq:wkb}
\end{equation}
Eq.~(\ref{eq:wkb}) implies that, in the absence of nonlinearities, the energy carried by Alfv\'en waves has a maximum at the critical point where the wind becomes super-Alfv\'enic, i.e., where $U=V_a$. Farther away from the Sun, where $U\sim U_0$ is nearly constant and $U_0\gg V_a$, the wave energy $u^2\propto R^{-1}$. Since, for an Alfv\'en wave, the fluctuating magnetic field ${\delta\bf B}$ is ${\delta\bf B}/\sqrt{4\pi\rho}=\pm {\bf u}$, Eq.~(\ref{eq:wkb}) implies that $\delta B^2\propto R^{-3}$. 
 An overall radial decrease of the root-mean-square (RMS) energies of fluctuations $\langle b^2\rangle$ has been reported by analyzing Helios data at radial distances $0.3\lesssim R\lesssim 1$~au \citep{bavassano1982radial, roberts1990amplitudes}. Not surprisingly, the low-frequency part of the Alfvénic spectrum was found to evolve in good agreement with the WKB theory, $\langle\delta B^2\rangle\propto R^{-3}$. However, higher frequencies were found to fall off with radial distance faster than predicted. In general, deviations from the WKB prediction are an indication of other effects, primarily the nonlinear cascade of energy, even though instabilities such as parametric decay could also come into play and affect the radial evolution of fluctuations (see, e.g., Sec. \ref{sec:pdi}), as well as large-scale gradients in the underlying plasma \citep{shi2020propagation,roberts1992velocity}. 
 
\begin{figure}
    \centering
    \includegraphics[width=0.45\textwidth]{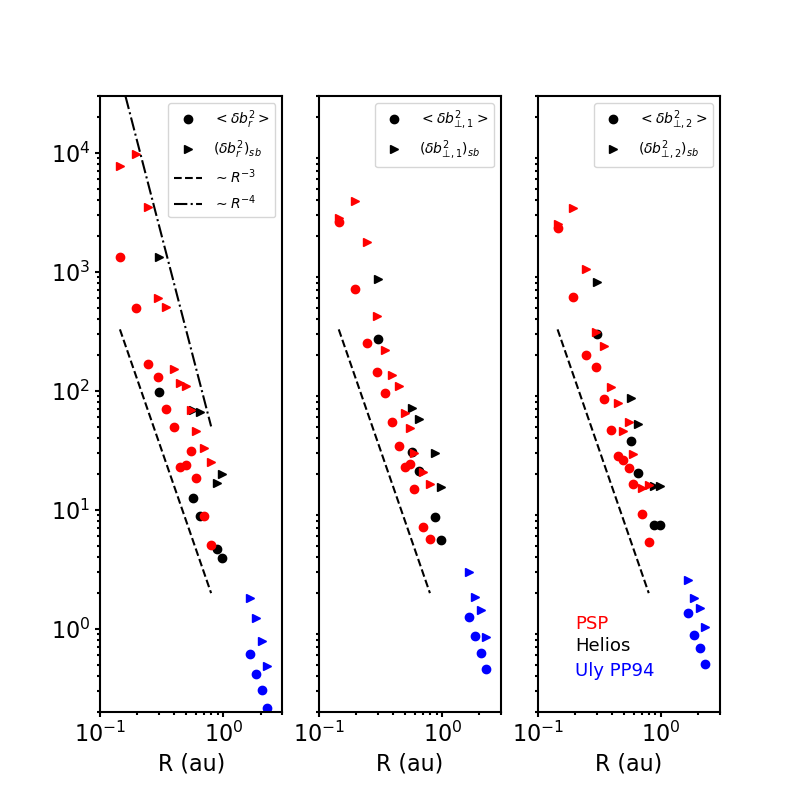}
    \includegraphics[width=0.45\textwidth]{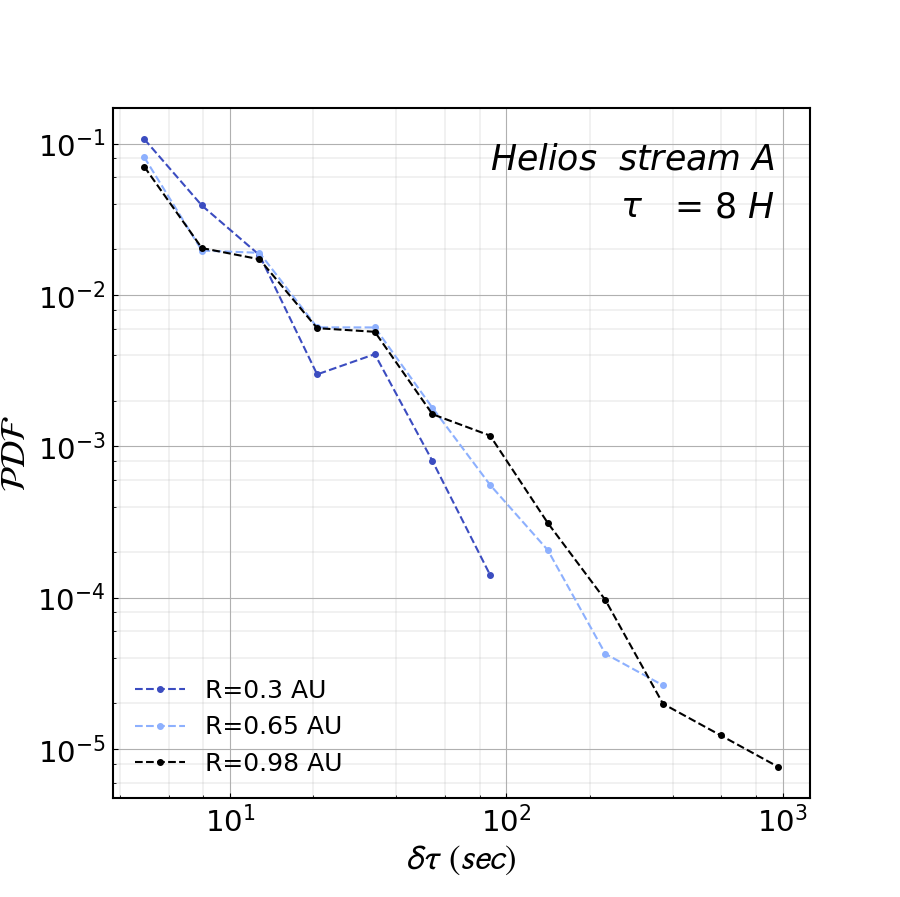}
    \caption{Left three panels: example of radial evolution of the rms of the energy  fluctuations where the dots correspond to the rms of energy fluctuations (radial, and two perpendicular directions from left to right), and the triangles correspond to the rms calculated over the sparse set of switchbacks. Right panel: probability distribution function of switchback duration $\delta\tau$ at different radial distances from Helios data. Figures reproduced with permission from \cite{Tenerani2021}, copyrights by AAS}
    \label{fig:rms_all}
\end{figure}

 \citet{Tenerani2021} performed a similar analysis by combining data from PSP, Helios, and Ulysses. In addition to recovering past results from Helios by  including data closer to the Sun (down to $R\simeq 0.2$~au), two new main results are worth highlighting. First, not only does the total RMS fluctuation energy follow the WKB predictions at low frequencies, but also the radial component of the magnetic field scales close to $\langle\delta B_r^2\rangle\propto R^{-3}$. 
 This implies that field-aligned fluctuations decay slower than the background magnetic field, supporting the scenario that expansion can also be a driver for the generation of switchbacks \citep{paper5}. 
 Second, the sparse ensemble of fluctuations corresponding to the largest amplitude switchbacks displays a decay rate with radial distance close to $\langle\delta B_r^2\rangle_{sb}\propto R^{-4}$, larger than the WKB one, even at the lowest frequencies. Figure \ref{fig:rms_all}, left panel, shows an example of radial evolution of the fluctuations' energy rms per component (from left to right, for the radial and two perpendicular components), where the dots correspond to the whole ensemble of fluctuations and the triangles correspond to the sparse set of switchbacks. It has been suggested \citep{Tenerani2021,matteini2024alfvenic} that the observed trends in the radial component of the largest amplitudes is a result of the waves' spherical polarization, or constant magnetic-field strength -- a property that is commonly observed in Alfv\'enic fluctuations in the solar wind, including switchbacks. The condition of spherical polarization imposes constraints on the relative amplitude of fluctuations because the total magnetic field strength,
 \begin{equation}
     B^2 = B_0^2 + \delta B^2 + 2\boldsymbol{B}_0\cdot\delta \boldsymbol{B}\label{eq:B2}
 \end{equation} 
 must remain locally constant over a fluctuation (while varying slowly with radial distance from the Sun).
 Writing the normalized RMS fluctuation amplitude 
$A={\delta B}/{B_0}$,
upon requiring $B^2$ constant, one finds for the parallel component
\begin{equation}
    \frac{\delta B_\parallel}{B_0} \sim \min(A^2, A\sin\theta),\label{eq:parfluc}
\end{equation}
where $\theta$ is {the angle between the direction in which the fluctuations vary (the direction of the gradients of $\delta \boldsymbol{B}$) and the mean magnetic field $\boldsymbol{B}_0$}. For a magnetic field reversal (a ``full switchback'') to occur, $\delta B_\parallel/B_0>1$. Because Alfv\'enic fluctuations are transverse to the wave's gradients ($\nabla \cdot \boldsymbol{B}=0$), reversals of the parallel field are easier for more perpendicular gradients. While this analysis formally applies only to one-dimensional waves, Equation~(\ref{eq:parfluc}) appears to be well-satisfied even in three-dimensional, turbulent simulations \citep{Johnston2022}.
 
 Recent simulations of two-dimensional turbulence using the Hybrid Expanding Box model provided compelling results showing that magnetic field fluctuations evolve towards  spherical polarisation \citep{matteini2024alfvenic}. This is achieved by the generation of the radial (parallel, since $\boldsymbol{B}_0$ is taken radial) fluctuations that are required to make $B^2$ constant, a process that has also been associated with forcing due to the small magnetic pressure fluctuations introduced as a result of expansion \citep{mallet2021evolution, barnes1974large}. To maintain spherical polarization close to the Sun, in the phase when $\delta B/B_0$ is growing, the radial fluctuating magnetic field must decay slower than the transverse fluctuations. Other simulations have similarly demonstrated that nearly constant-$B$ fluctuations leading to switchbacks are generated 
in a turbulent expanding medium \citep{Squire2020,Shoda2021}. Even though the properties of switchbacks can be characterized from their geometry, it remains to be understood what the physical process(es) leading to spherical polarization are and if the observed non-WKB radial evolution of the largest  switchbacks has a dynamic origin. Tracking how switchbacks evolve as the solar wind expands may provide useful insights to answer to the latter question, as discussed in the next Section.

\subsection{Observations of Switchbacks' Occurrence}\label{sec:observation_occurrence}

 The results summarized in Section \ref{sec:amplitudeevo} indicate that solar wind expansion can act as a driver for \emph{in situ} switchback generation. On the other hand, theory and simulations also show that switchbacks (and in general large amplitude Alfvén waves) eventually become unstable \citep{Tenerani2020,shi2024analytic,marriott2024parametric}, suffer strong reflection when gradients are present \citep{Magyar2021b}, or may simply undergo dispersion \citep{mallet2023nonlinear,tenerani2023dispersive} by emitting waves at their boundaries, resulting in a reduction of wave energy. Some of these processes are described in subsequent sections, including observations consistent with switchback boundary erosion by magnetic field reconnection. Arguably, if any of those processes are at play, then one should expect a competition between wave decay/erosion and expansion that might leave signatures on the occurrence of switchbacks with radial distance. 
 
 A series of works has investigated how the occurrence rate of switchbacks depends on the radial distance. \cite{Mozer2021} determined the occurrence rate of switchbacks, defined as deflections of the magnetic field larger than $90^\circ$ from the Parker spiral, and found that the switchback occurrence rate is independent of radial distance. This result differs from subsequent studies using larger datasets and where switchbacks were defined as $90^\circ$ rotation of the magnetic field to the mean field, which found a radial dependence of the occurrence rate, as well as a dependence on the scale of the switchbacks \citep{Tenerani2021,Jagarlamudi2023}. 
The right panel of Fig.~\ref{fig:rms_all} shows the probability distribution of switchbacks as a function of their duration at different radial distances from Helios. It demonstrates that the probability of finding longer duration switchbacks increases with radial distance, while it decreases for shorter duration switchbacks. Similar trends are also found at Parker Solar Probe and Ulysses.
\citet{Jagarlamudi2023} analyze the evolution of the occurrence of switchbacks based on data from the Parker Solar Probe's first 10 encounters between 13.28 and 58 solar radii. They find that the percentage of the number of short-duration switchbacks decreases with distance, while the long-duration switchbacks increase, consistent with the results of \citet{Tenerani2021}.
These results suggest that the evolution of switchbacks is scale-dependent, and that both the \emph{in situ} generation and decay are at play simultaneously. 
There is also evidence that the occurrence rate of switchbacks is reduced inside the Alfv\'en critical point more than outside \citep{Pecora2022,2024ApJ...970L..26A}. 
If this is true, it implies that most of the switchbacks are generated \emph{in situ} instead of in the lower corona.
However, there are only a few observations from inside the Alfv\'en radius compared to from larger radial distances. 
As PSP continues lowering its orbit and collects more data inside the Alfv\'en radius, a further study is warranted to draw statistically relevant conclusions on \emph{in situ} versus coronal origins.

\begin{figure}[htb!]
    \centering
    \includegraphics[width=\textwidth]{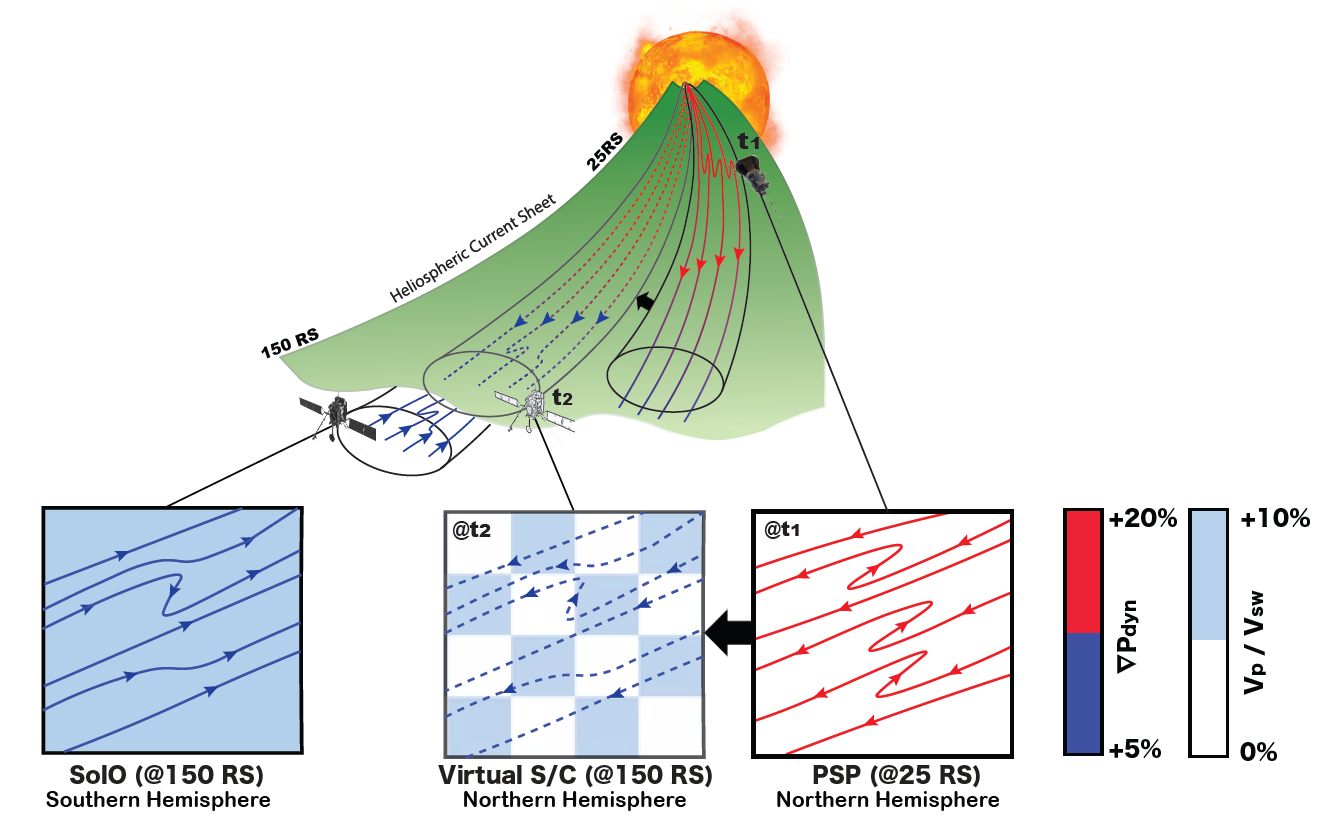}
    \caption{Concept illustration of the spatial and temporal evolution of a magnetic
switchback. The color bars indicate dynamic pressure ($P_{dyn}$) and relative velocity
($V_p/V_{sw}$). Figure reproduced with permission from \citep{Soni2024arXiv240213964L}, copyright by AAS}
    \label{fig:PSPSOLO_fig4}
\end{figure}

As first pointed out by \citet{Horbury2020}, magnetic switchbacks usually appear in ``patches''. Each individual switchback can last for minutes, but their occurrence, as well as amplitude, often display a larger-period 
{modulation}, from tens of minutes to several hours.
Throughout the duration of a switchback patch, the average solar wind parameters, including magnetic field strength, density, alpha particle abundance, and solar wind speed, vary slowly \citep{Bale2021,Shi2022}.
Although whether the switchback patches are generated by spatial structures, e.g. magnetic funnels on top of the supergranule networks \citep{Bale2021,Fargette2021}, or temporal processes, such as periodic breathing of emerging magnetic fluxes \citep{Shi2022}, is still unclear, the evolution of these patches may significantly modify the meso-scale solar wind parameters.
In a recent work, \citet{Soni2024arXiv240213964L} compare one interval of PSP observation with an interval of Solar Orbiter (SO) observation. They suggest that a switchback patch near the Sun may gradually relax into a ``microstream'' as it propagates, releasing energy to the background solar wind. 
During this relaxation process, the number of switchbacks inside the patch declines by $\sim 30$\% while the background proton velocity is enhanced and becomes 10\% greater than the pristine solar wind. 
A conceptual illustration of the spatial and temporal evolution of a magnetic switchback patch is presented in Figure \ref{fig:PSPSOLO_fig4}.
However, we point out that, in \citet{Soni2024arXiv240213964L}, the stream observed by PSP is not necessarily the same stream observed by SO.
Therefore, to solidify their conclusions, a future statistical analysis of the properties of switchback patches at different radial locations is necessary.

\subsection{Parker Spiral Effects}\label{sec:parkerspiral}
The expansion-driven evolution of switchback amplitudes outlined in Section~\ref{sec:amplitudeevo} must be modified to include the effects of the Parker spiral. It is convenient to adopt the radial-transverse-normal (RTN) coordinates\footnote{$\hat{\boldsymbol{e}}_R$ is the outward radial direction from the center of the Sun to the observer. $\hat{\boldsymbol{e}}_T$ is the cross product of the northward solar rotation axis and $\hat{\boldsymbol{e}}_R$. $\hat{\boldsymbol{e}}_N=\hat{\boldsymbol{e}}_R \times \hat{\boldsymbol{e}}_T$.} in describing the solar magnetic field. Due to solar rotation, far from the Sun, the mean radial magnetic field obeys $B_R \propto R^{-2}$, while the mean tangential component obeys $B_T \propto R^{-1}$ \citep{parker1958dynamics}, dragging the field lines into an Archimedean spiral (the ``Parker spiral"). \cite{Squire2022} and \cite{Johnston2022} have studied the effect of the Parker spiral on the growth of Alfv\'enic switchbacks analytically and numerically, with a number of important conclusions: the spiral field enhances the \emph{in situ} growth of switchbacks and introduces significant asymmetries into the switchback structure. This is an important example of how a detailed understanding of switchback evolution must be taken into account for assessing their origin, and in interpreting asymmetries present in the data \citep{Fargette2021}, which may be the result of \emph{in situ} evolution rather than necessarily the imprint of the process at the origin.

The effects of the Parker spiral on switchbacks may be understood in terms of the interplay of several different physical effects. 
First, the growth of the normalized amplitude $A$ is eventually `arrested' by the Parker spiral: while for a constant solar wind velocity, {the root-mean-square magnetic fluctuation amplitude $\delta B_{rms} \propto R^{-3/2}$} due to wave-action conservation as shown before, the mean field $\boldsymbol{B}_0 = \hat{\boldsymbol{e}}_R B_R + \hat{\boldsymbol{e}}_T B_T$, with $B_R \propto R^{-2}$ but $B_T \propto R^{-1}$. Thus,
{\begin{equation}
A = \frac{\delta B_{rms}}{B_0} = \frac{A_0 (R/R_0)^{1/2}}{\sqrt{1+(B_{T0}/B_{R0})^2(R/R_0)^2}},\label{eq:Aps}
\end{equation}}
where $B_{R0}$ and $B_{T0}$ are the radial and transverse components of the mean Parker spiral field at some radius $R_0$: for $R/R_0 \ll B_{T0}/B_{R0}$, the field is mainly radial and $A \sim (R/R_0)^{1/2}$, but for $R/R_0 \gg B_{T0}/B_{R0}$, the field has spiralled into the tangential direction, and $A \sim (R/R_0)^{-1/2}$, decreasing with increasing radius, and thus potentially decreasing the number of switchbacks. However, this point occurs far out compared to the distances studied by PSP.

Second, as the solar wind travels out from the Sun, it expands in the transverse directions, and the gradients of the waves embedded in it thus rotate towards the radial direction{: as the plasma expands in the transverse directionsas the wind propagates from $R_0$ to $R$,
\begin{equation}\label{eq:gradient_operator_with_expansion}
    \nabla \to (\partial_R,(R_0/R)\partial_T,(R_0/R)\partial_N),
\end{equation}
so a vector along the direction of this gradient, $\boldsymbol{p}$, say, rotates towards the radial direction.} At the same time, the Parker spiral mean field direction is rotating away from the radial direction {as described above:}
this gives rise to a complex non-monotonic behaviour of $\theta$, the angle of the wave gradients to the mean field direction{, expressed mathematically as
\begin{equation}
    \sin^2\theta = 1- \frac{(\boldsymbol{p}\cdot \boldsymbol{B}_0)^2}{|p|^2|B_0|^2}.
\end{equation}
I}nitially, $\boldsymbol{B}_0$ is radial enough that $\theta$ decreases, but at a distance
\begin{equation}
\frac{R_{\rm min}}{R_0} = \sqrt{\cot\theta_{\perp0} \cot\Phi_0},
\end{equation}
where $\theta_{\perp0}$ is the angle the direction of the gradients makes to the $T-N$ plane at $R_0$ and $\Phi_0=\arctan(B_{R0}/B_{T0})$ is the Parker spiral angle at $R_0$, $\theta$ reaches a local minimum and starts to increase again. Using Equation~(\ref{eq:parfluc}), once $A$ is large enough that $\delta B_\parallel$ also depends on $\sin\theta$, beyond $R_{\rm min}$ this effect tends to promote the generation of switchbacks.
\begin{figure}
    \centering
    \includegraphics[width=0.5\linewidth]{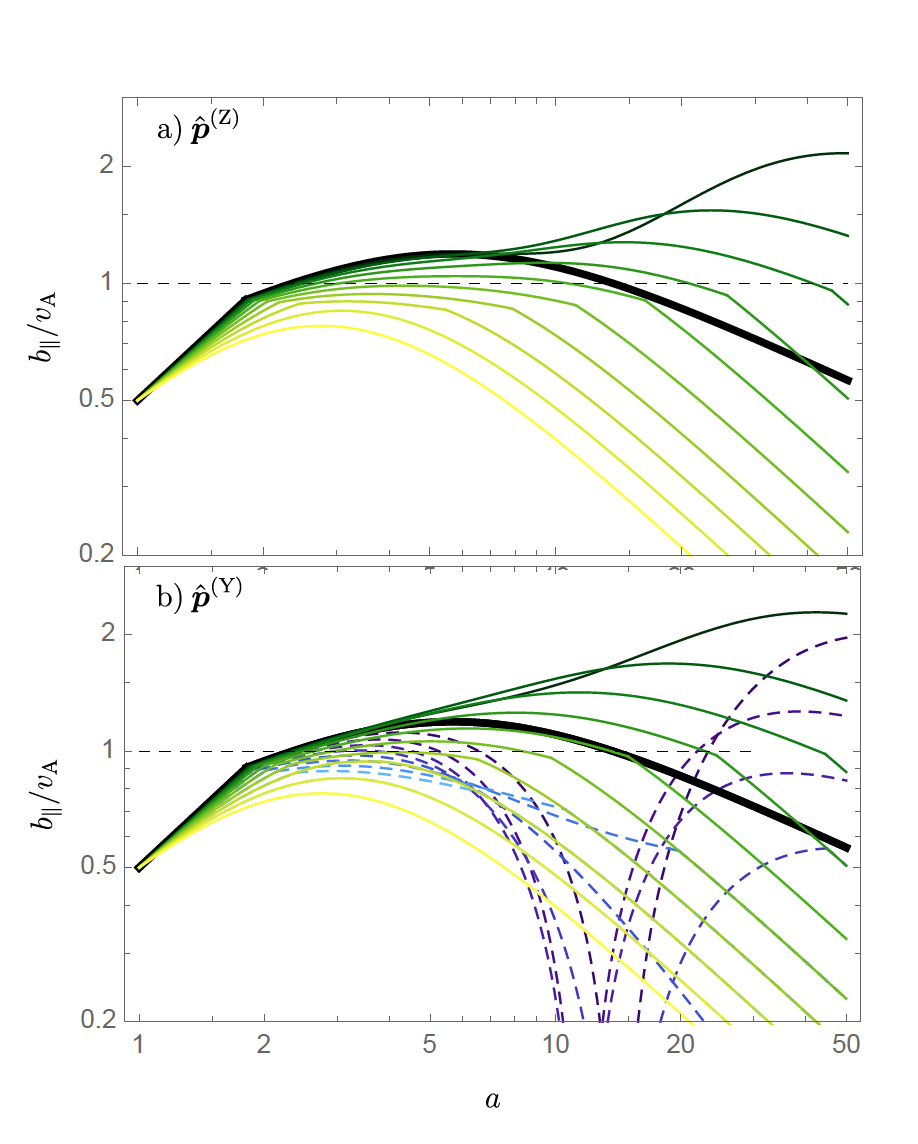}
    \caption{Evolution of the parallel magnetic-field fluctuation, roughly equivalent to the switchback prevalence, computed from Eq.~(\ref{eq:parfluc}) using the scalings for the amplitude (Eq.~\ref{eq:Aps}) and gradient angle to parallel $\theta$ implied by the expanding box model with the Parker spiral. Horizontal axis is the expansion parameter $a=R(t)/R(0)$ with $R(t)$ being the radial location of the expanding box. Thick black lines show the case with no Parker spiral, while coloured lines from dark green to yellow correspond to initial Parker spiral angles $\Phi_0=2^\circ,4^\circ,\ldots, 20^\circ$. The top panel shows the case where the gradients lie in the $R$-$N$ plane, while the bottom panel shows the case where the gradients lie in the $R$-$T$ plane of the Parker spiral. The blue dashed lines show the case where the gradient direction passes through parallel ($\theta=0$), at which point the parallel magnetic field fluctuation is zero. Reproduced with permission  from \cite{Squire2022}, copyright by AIP Publishing}
    \label{fig:ps_bpar}
\end{figure}
For $\theta_{\perp0},\Phi_0 \ll 1$, but $\theta_{\perp0} \gtrsim \Phi_0$, this minimum occurs at a smaller distance than the maximum in $A$ imposed by the Parker spiral: for such a situation, as $R$ increases, the amplitude $A$ is growing, and also $\theta$ is becoming more perpendicular, both of which promote switchback formation. This behaviour is illustrated in Figure~\ref{fig:ps_bpar}, where the black curves show the parallel magnetic-field fluctuation implied by Equation (\ref{eq:parfluc}) for a purely radial background magnetic field: eventually, $\theta$ rotates towards zero and the amplitude decays. In contrast, the green to yellow curves show increasing initial Parker spiral angles: as can be seen, there is a range over which a small Parker spiral dramatically increases the parallel magnetic field fluctuation, and thus the propensity to form switchbacks. This seems to be the relevant case in the solar wind, where waves have a range of $\theta_{\perp0}$ and the Parker spiral at small $R$ is nearly radial, with very small $\Phi_0$. Note that this relies on the waves having sufficiently large initial amplitudes that they reach the large-amplitude regime of Equation~{(\ref{eq:parfluc})}; roughly, $A_0 > \sqrt{\Phi_0}$ to form switchbacks, not a very restrictive bound.

As shown by many authors \citep{Horbury2020, Fargette2021, Laker2022}, switchbacks present a systematic preferential deflection along the tangential direction in RTN coordinates. The choice of the background magnetic field is crucial in assessing this property \citep{Fargette2022}. Since the full distribution of the magnetic field measurements has its peak at the local Parker spiral, such a frame seems the most physical motivated one \citep{Fargette2022}.

\begin{figure}
    \centering
    \includegraphics[width=1\textwidth]{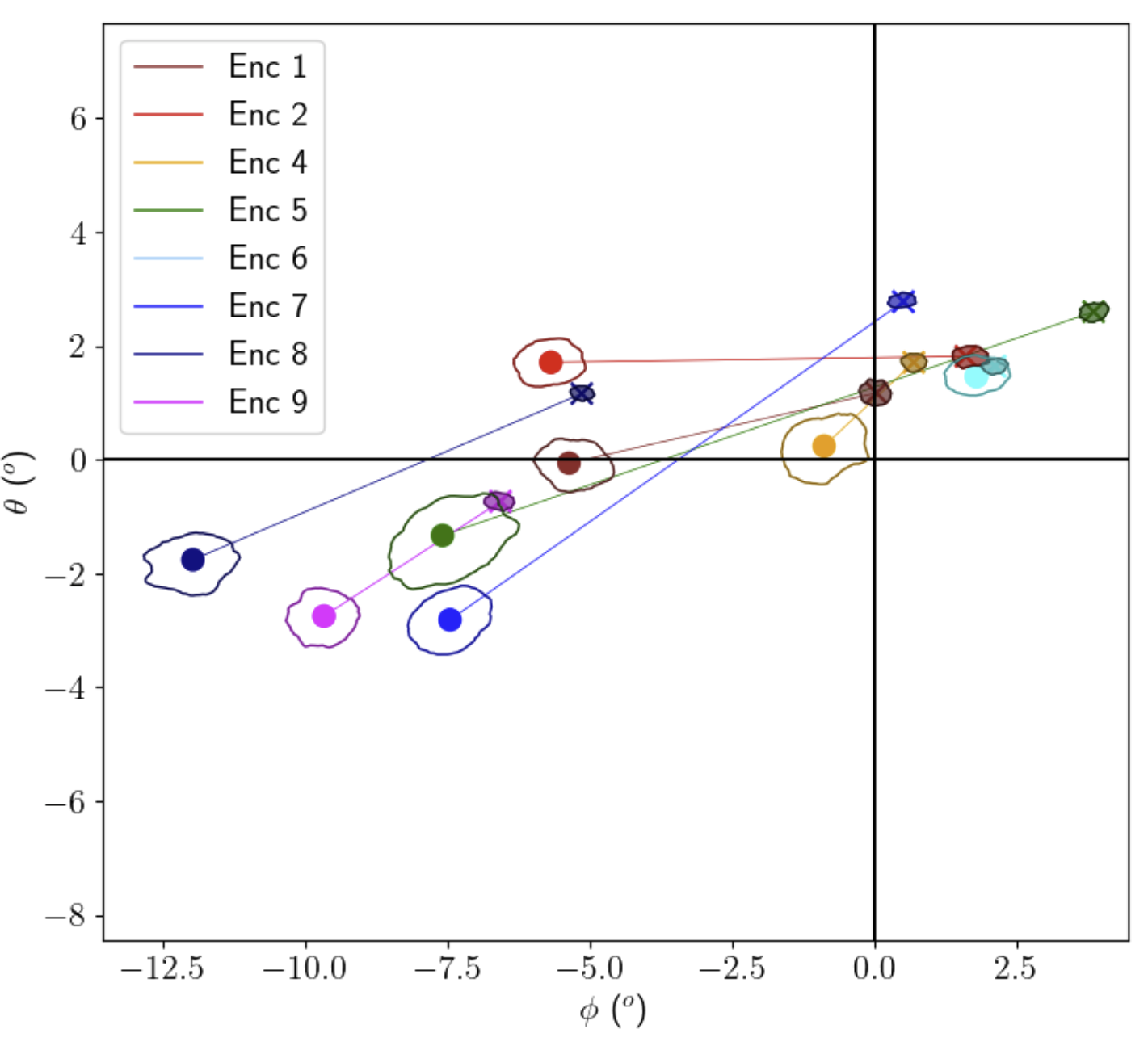}

    \caption{Deflections with respect to the Parker spiral in latitude $\theta$ and azimuth $\phi$. Mean values for the quiet solar wind (cross) and mean values for the switchback population (dot). Contours indicate the uncertainty of the performed fit. Figure reproduced with permission from \cite{Fargette2022}, copyright by ESO}
    \label{fig:fargette_deflections}
\end{figure}

The modeling of the measurements through a superposition of two Gaussian functions allows to fit well the magnetic field measurements with two populations, one that represents the quiet radial field (close to the Parker spiral) and another that represents the switchbacks population \citep{Fargette2022}. This approach reveals that the azimuth and latitude angles $[\phi, \theta]$ distributions for the switchback population (where $\phi=0$ and $\theta=0$ correspond to the local Parker spiral) show a systematic mean deflection towards negative values of the order of few degrees for $\phi$ and around zero for $\theta$ for encounters 1 to 9 (except encounter 6).
The rotation in $\phi$ translates to a deflection skewed towards the $+ \hat{\boldsymbol{e}}_T$ direction if the field has a negative polarity or $-\hat{\boldsymbol{e}}_T$ if the polarity is positive. A summary of these results is shown in Figure~\ref{fig:fargette_deflections}. The existence of a preferential deflection direction is also observed in  \cite{Laker2022} with a different technique. The authors look at the distributions of the clock angle (the angle of the magnetic field in the plane normal to the local Parker spiral). Their findings confirm that overall the preferential deflection is along the T directions but their results are not in agreement with the results of \cite{Fargette2022} for every encounter. In fact the association negative (positive) polarity $+(-) \hat{\boldsymbol{e}}_T$ direction is not observed in the clock angle for all the encounters. 
For example, for Encounter 4 \citep[see Figure~4 of][]{Laker2022} the clock angle distribution relative to the negative polarity is more skewed towards the -T direction. The origin of these differences is probably due to the different techniques used and to the always present stream-to-stream variability that is surely not captured in the mean of the switchbacks fitted distribution used in \citep{Fargette2022}.

Perhaps surprisingly, this preferential deflection can arise simply from the geometrical constraints required by the Alfv\'enic nature of the switchbacks \citep{Squire2022}. Including the Parker spiral, the {instantaneous} parallel magnetic field fluctuation that drives a switchback has both radial and tangential components, $\delta \boldsymbol{B}_\parallel = \delta B_R \hat{\boldsymbol{e}}_R + \delta B_T \hat{\boldsymbol{e}}_T$: for very smaller Parker spiral angles, switchback fluctuations must be dominated by $\delta B_R$, but \cite{Squire2022} show that beyond $R_{\rm min}$, they are instead dominated by $\delta B_T$. Thus, switchbacks are associated with places where the total tangential magnetic field $B_T = \delta B_T + B_{0T}$ passes through zero. Because the total magnetic field strength must remain constant (since we have assumed that the switchback is Alfv\'enic), this means that the total radial magnetic field must increase to compensate: the direction of the radial magnetic field fluctuation is always in the direction of the background radial mean field. In other words, due to the constant magnetic-field-strength condition, the magnetic field vector in switchbacks should preferentially rotate towards the radial direction from the background Parker spiral direction, as seen in the data \citep{Horbury2020,DudokdeWit2020,Laker2022} and in numerical simulations \citep{Johnston2022}.

\subsection{Effect of Expansion on Switchback Boundaries}\label{sec:boundaries}

\subsubsection{Theoretical Modelling of Discontinuity Formation}

Switchbacks are often observed to have remarkably steep, discontinuous boundaries. As is true for multiple aspects of switchback physics, there are two main possibilities for the physical origin of the boundary discontinuities: they are born already as discontinuities deep in the corona and then evolve as they travel outwards, or, alternatively, they form \emph{in situ} via a steepening process. We will focus on the latter possibility first, returning to the non-ideal evolution of the boundaries in Sec.~\ref{sec:nonideal_boundaries}.

As described in Sec.~\ref{sec:amplitudeevo}, in an inhomogeneous solar wind with a radial magnetic field, the normalized amplitude $A=\delta B/B$ of Alfv\'en waves naturally increases with distance from the Sun. \citet{Mallet2021} show that once the waves reach large amplitude $\delta B/B\approx 1$, there is also significant steepening as a part of this expansion-driven evolution. The physical origin of this steepening is as follows. As explained in Sec.~\ref{sec:amplitudeevo}, as the solar wind (which we assume for simplicity has constant velocity) expands outwards, Alfv\'en waves have amplitudes $\delta B \sim R^{-3/2}$, while the background magnetic field (which we assume here to be radial for simplicity) behaves as $B_0 \propto R^{-2}$. Let us assume the wave varies on short length scales, over which $B_0$ is locally constant in the $x$ direction. The total magnetic field strength, however,
is then {linearly} not constant over the short lengthscales over which the fluctuations vary{: rewriting Eq.~(\ref{eq:B2}),
\begin{equation}
    B^2 = B_0^2 + \delta B^2 + 2\delta \boldsymbol{B}\cdot\boldsymbol{B_0},
\end{equation}
{linearly} all three components in the expression above have different scalings with $R$, with the first two terms scaling as $R^{-4}$ and $R^{-3}$ respectively\footnote{The \emph{linear} scaling of the final term depends on the angle between the local gradient direction of the fluctuation and the background magnetic field: since this argument is for illustrative purposes only, we will not expand on this further here, but readers may consult \citet{mallet2021evolution}, Eq. 59 to derive the precise linear scaling.}.} Thus, {without taking the nonlinearity into account,} expansion is constantly generating fluctuations in the magnetic field strength. {Nonlinearly,} this fluctuation in magnetic pressure drives a compressive flow, which distorts and steepens the wave, opposing the generation of the magnetic pressure that created it{, and dramatically reducing the fluctuations in $B^2$ compared to the linear supposition implied by the linear WKB scalings.} This picture explains how switchbacks maintain approximately constant magnetic-field magnitude as they grow, while also providing a physical process by which their boundaries steepen.
\begin{figure}
    \centering
    \includegraphics[width=\linewidth]{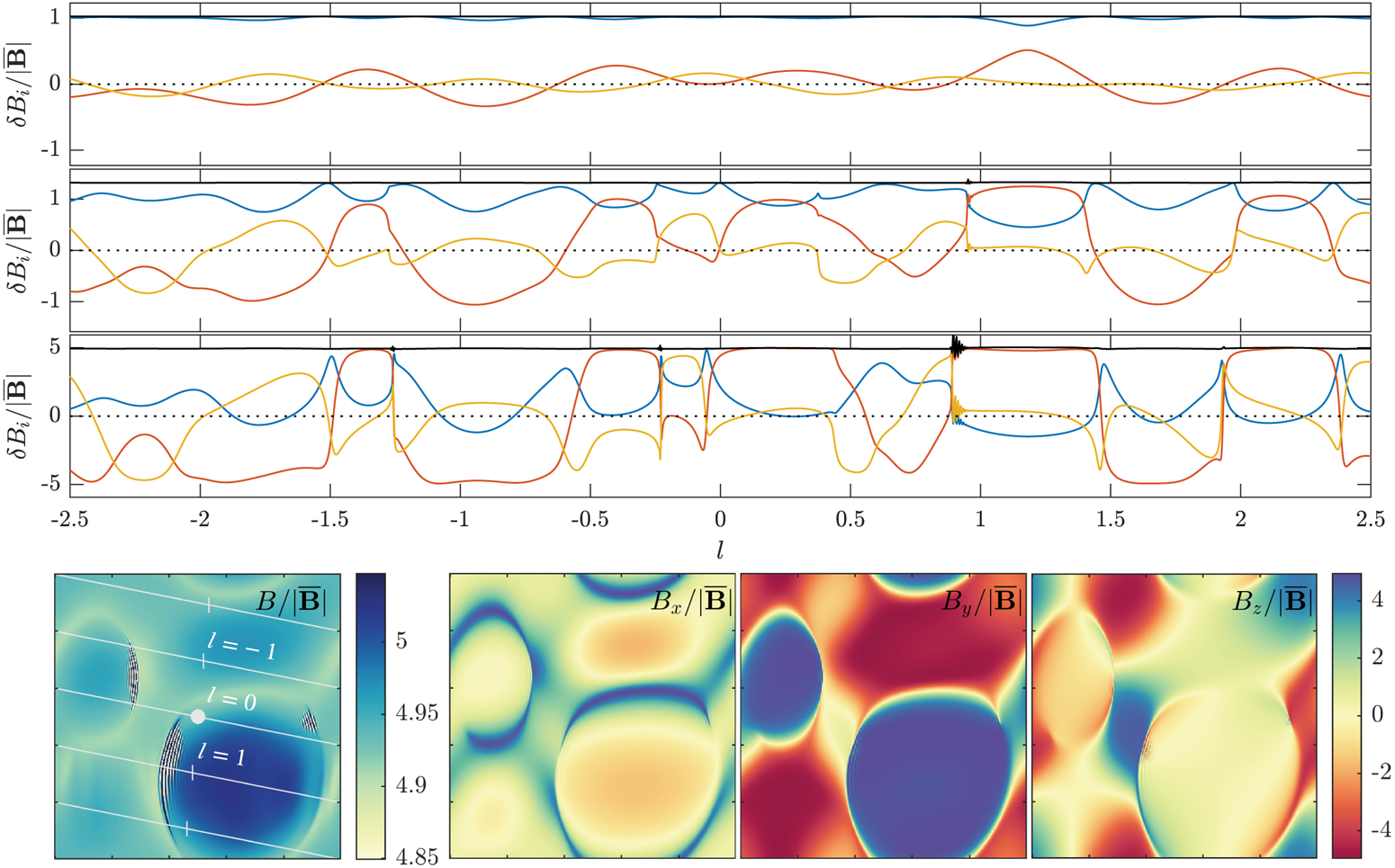}
    \caption{Two-dimensional spherically-polarized solution on a plane angled at $\theta=30^\circ$ from the mean-field direction. The top three panels show periodic traces of $|\boldsymbol{B}|$ (black), $B_x$ (blue), $B_y$ (red), and $B_z$ (yellow), along the white line plotted on the bottom-left panel. The initial condition was constructed from a random superposition of modes in $A_x$, scaled to give an initial amplitude $A\approx 0.2$ (top panel). The solution then grows in time from an exponential term added to the induction equation (see text), with time (and amplitude) increasing from the first to the third traces. The bottom panels show the 2-D structure of the components at the time corresponding to the bottom trace, when the amplitude $A\approx 5$. Discontinuities appear naturally in the solution as a result of the amplitude growth: however, apart from numerical ringing, $|\boldsymbol{B}|$ remains constant to within 2\% throughout the domain. Figure reproduced with permission from \cite{squire2022b}, copyright by AIP Publishing}
    \label{fig:squire2022steep}
\end{figure}
However, in the simplified one-dimensional model considered by \citet{Mallet2021}, this only led to modest steepening. \citet{squire2022b} developed a heuristic model of three-dimensional MHD Alfv\'en waves, growing in amplitude in a similar way to the inhomogeneity-induced evolution in Sec.~\ref{sec:expansion}. Interestingly, their results show that some initial conditions do appear to produce true discontinuities (solutions whose steepness does not converge with increasing resolution), as shown in Fig.~\ref{fig:squire2022steep}. This is reminiscent of earlier results \citep{roberts2012,valentini2019} which showed that seeking a 3D divergenceless field that minimizes its strength leads to discontinuities.  Thus, it is possible that  sharp switchback boundaries may be another consequence of the physics of Alfv\'en wave growth in the expanding solar wind. Of course, how these discontinuous boundaries evolve once they form is an important and interesting question, which we will address in Sec.~\ref{sec:nonideal_boundaries}.

\subsubsection{Observations of Boundary Discontinuity Evolution}
\label{sec:boundaries_observations}
To discuss the observations of switchback boundary evolution, we first briefly recap the observational technique of characterizing boundaries, which is described in much more detail in \citet{paper3}. Of the plasma discontinuities, the two appropriate for switchback boundaries are \emph{tangential discontinuities} \citep[TDs, ][]{Hudson1970} , which are closed discontinuities with no plasma flow across the discontinuity, and \emph{rotational discontinuities} (RDs), which require (Alfv\'enic) flow across the boundary and constitute a pure rotation in the magnetic field with no change in the magnetic field strength. Identifying the vector normal to the plane of the discontinuity is required for the classification of the boundary. 
Usually, the classification of observed discontinuities is then based based on two parameters \citep{Smith1973}: the degree of collinearity $(\textbf{\textit{B}}/|\textbf{\textit{B}}|) \cdot \textbf{\textit{n}}$, where $\bm{n}$ is the unit vector normal to the discontinuity, is compared to the jump in field magnitude across the discontinuity $\Delta |\textbf{\textit{B}}|/|\textbf{\textit{B}}|$. A threshold is needed to distinguish the categories: for example, \citet{Neugebauer1984} (and many subsequent works, including this review) select 0.2 for the jump and 0.4 for the collinearity, see Table \ref{tab:disc}. Note that these classification scheme does not include any details of the flow: considering that this is one of the defining physical differences between the tangential and rotational discontinuities, any statistical results should be viewed with caution. Of particular relevance to switchbacks, a steepened nonlinear MHD Alfv\'en wave, which necessarily has $\Delta |\textbf{\textit{B}}|/|\textbf{\textit{B}}|=0$, would be classified as either ``rotational" or ``either", dependent on the angle of propagation $\theta$ with respect to the mean field: the normal component being $B_0\cos\theta$.

\begin{table}[]
\begin{tabular}{c|c|c|c|c}
\caption{Thresholds for the classification of boundary discontinuities.}
Type                                                                    & Rotational {(RD)} & Tangential {(TD)} & Either {(ED)} & Neither {(ND)}    \\
\hline
\hline
$(\textbf{\textit{B}}/|\textbf{\textit{B}}|) \cdot \textbf{\textit{n}}$ & $\geq 0.4$ & $<0.4$     & $<0.4$ & $\geq 0.4$ \\
\hline
$\Delta |\textbf{\textit{B}}|/|\textbf{\textit{B}}|$                    & $<0.2$     & $\geq 0.2$ & $<0.2$ & $\geq 0.2$
\end{tabular}
\vskip 12 pt
\label{tab:disc}
\end{table}

Moreover, identifying the normal direction is non-trivial considering the single cut through the plasma provided by spacecraft observations. A recurrent methodology found in the literature is to apply the Minimum Variance Analysis \citep[MVA;][]{Sonnerup1998} to identify the plane of the discontinuity. From that plane, the normal component of the magnetic field can be computed and used to distinguish the nature of the boundary. This methodology is used in recent studies of the nature of switchback boundaries \citep{Larosa2021,AkhavanTafti2021,AkhavanTafti2022}. However, a recent study showed the limits of the MVA to correctly identify most of the planes in the context of switchbacks boundaries \citep{Bizien2023}. The reasons stated are the Alfvénicity of the switchbacks, which limits the jumps of magnetic field magnitude, as well as the superimposed fluctuations, causing an immediate ill-defined plane of the discontinuity. They propose a new methodology based on MVA and Singular Value Decomposition \citep[SVD,][]{Golub2013}, which can correctly identify most of the boundaries. More details of the methodology and its limitations, as well as the statistics of switchback boundaries, may be found in \citet{paper3}.

Considering the reasons stated above, the analysis of the evolution of the switchback boundary nature should be taken with precautions when comparing different studies. \citet{Larosa2021}, \citet{AkhavanTafti2021} and \citet{Bizien2023} focused on the first encounter of PSP, and thus do not directly address the evolution of boundary type. The results based on MVA alone show similar results: \citet{Larosa2021} finds more RDs than TDs but a large proportion are in the ED category, while \citet{AkhavanTafti2021} finds that the majority of the boundaries are RDs. However, \citet{Bizien2023} presents results based on the combined use of SVD and MVA, which reduce the bias towards RDs and thus finds mostly TDs.

\begin{figure}
    \centering
    \includegraphics{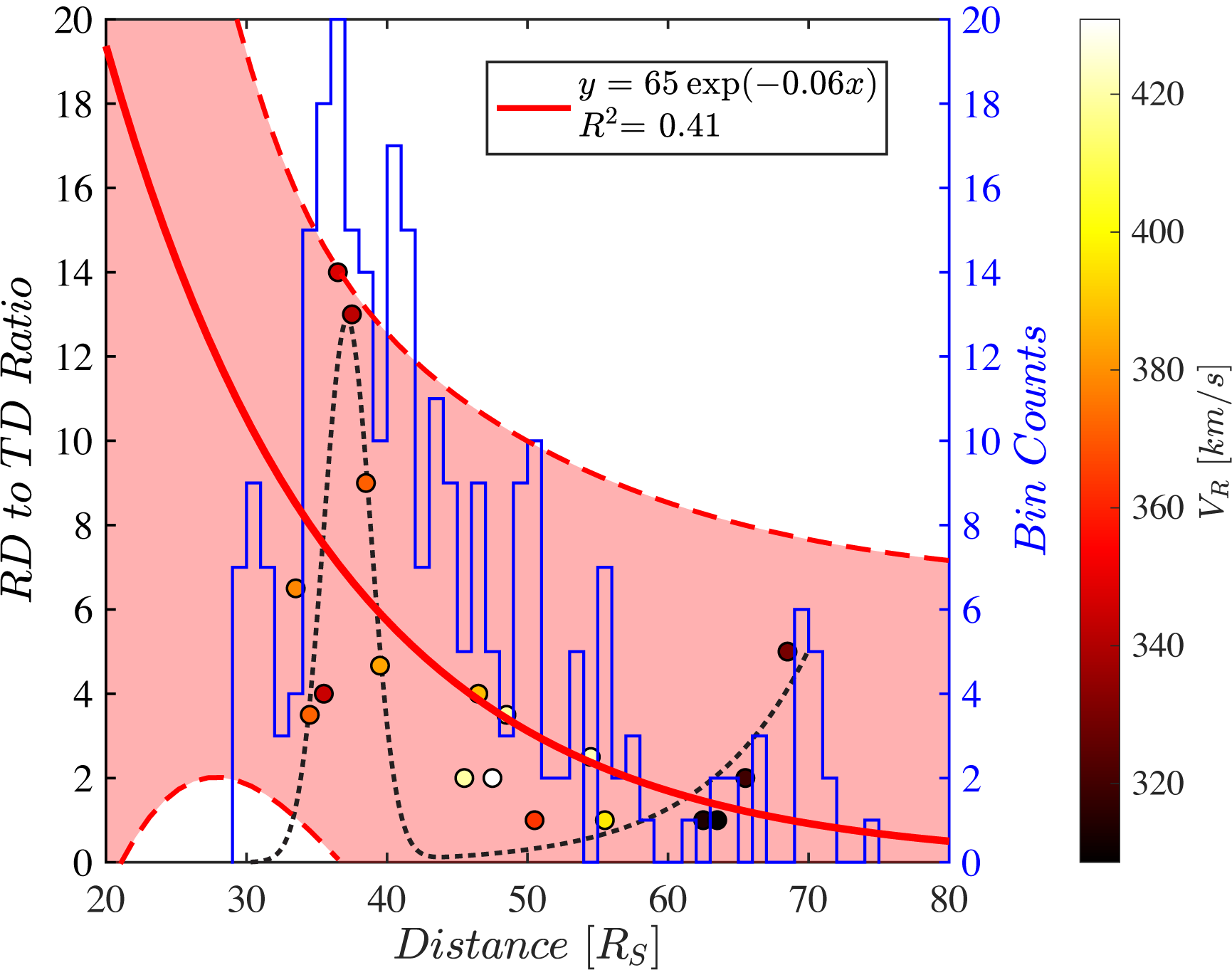}
    \caption{Plot of the RD-to-TD event ratio as a function of heliocentric distance at leading switchback transition regions. Color bar represents the average proton velocity in the leading quiet region before the switchback. The blue histogram represents the event counts inside each $1R_{\odot}$-wide distance bin. The solid red curve represents an exponential fit, as shown in the legend. The dashed black line represents a two-term Gaussian fit to the RD-to-TD ratio. The red shaded region represents the 95\% confidence interval for the RD-to-TD ratio. Figure reproduced with permission from \cite{AkhavanTafti2022}, copyright by AAS}
    \label{fig:akhavantafti2022_3b}
\end{figure}

\citet{AkhavanTafti2022} analyzed the first eight encounters and therefore showed more emphasis on the radial evolution. Their results show an exponential decay of the ratio of rotational to tangential discontinuities with respect to the radial distance from the Sun, and suggest that this evolution may be accompanied by the dissipation of the switchbacks' magnetic energy. However, the evolution of the switchbacks in the inhomogeneous solar wind under conservation of wave action also needs to be taken into account (see Sec.\ref{sec:amplitudeevo} and \ref{sec:parkerspiral}): for a radial background field, {the normal direction to the discontinuity naturally rotates toward the mean field because of the spherical expansion of the solar wind (Equation (\ref{eq:gradient_operator_with_expansion})) while the (absolute) amplitude of Alfv\'en waves decreases with radial distance}. Moreover, these conclusions should now be reconsidered as the identified rotational discontinuities may need to be reclassified in light of the methodology proposed by \citet{Bizien2023}.

\subsection{Radial Evolution of Alfv\'enicity}\label{sec:alfvenicity}

\begin{figure}
    \centering
    \includegraphics[width=\hsize]{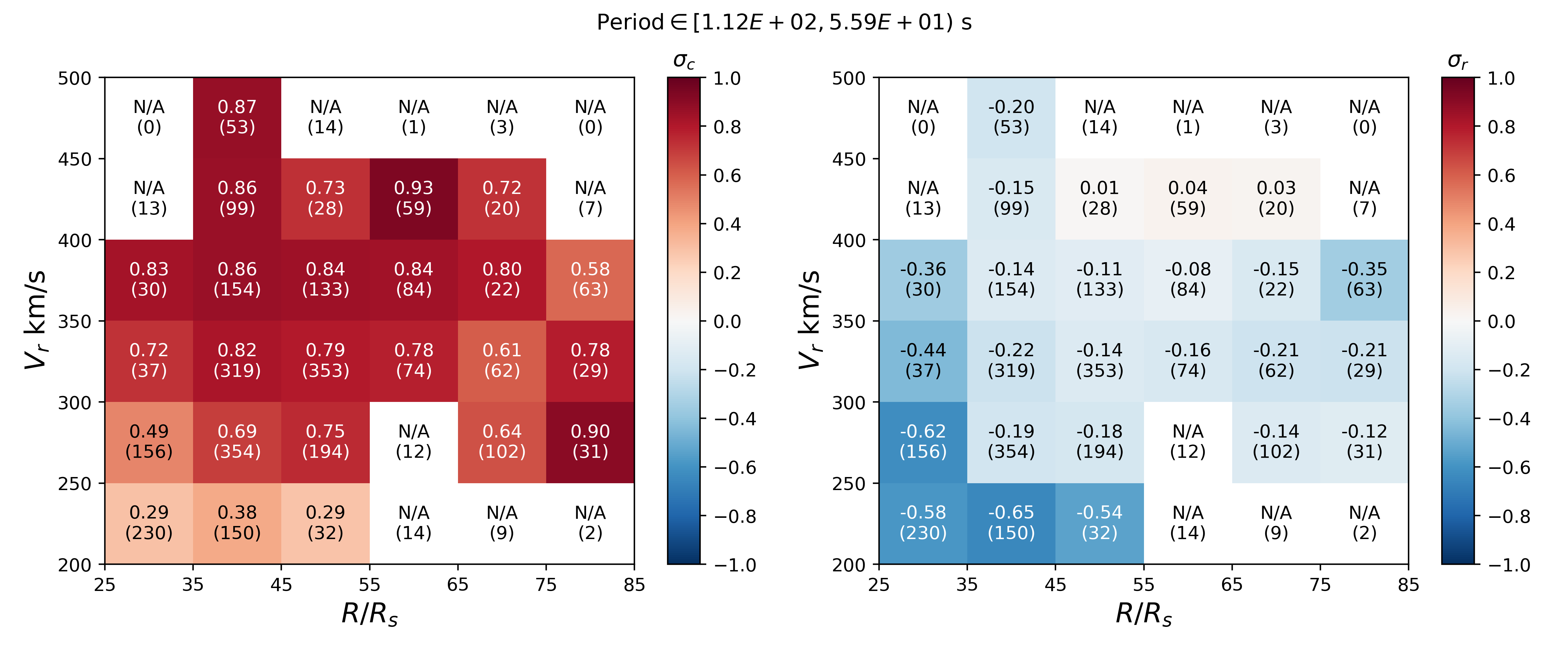}
    \caption{Normalized cross helicity $\sigma_c$ (left) and normalized residual energy $\sigma_r$ (right) as functions of the radial distance to the Sun $R$ and the radial speed of solar wind $V_r$. The colors of each block represent the median values of the binned data. Text on each block shows the value of the block and the number of data points (in brackets) in the block. Bins with no more than 15 data points were discarded. The results are calculated for wave periods $112s \geq T \geq 56s$. Figure reproduced with permission from \citet{Shi2021}, copyright by EOS}
    \label{fig:shi2021_fig08}
\end{figure}

Switchbacks are mostly Alfv\'enic structures embedded in the solar wind turbulence. Thus, there is no doubt that the evolution of the Alfv\'enicity of the solar wind turbulence is highly related to the evolution of switchbacks.
The most widely used diagnostics of the Alfv\'enicity of MHD turbulence are normalized cross helicity 
\begin{equation}
    \sigma_c = \frac{\left| \bm{z^+} \right|^2 - \left| \bm{z^-} \right|^2}{\left| \bm{z^+} \right|^2 + \left| \bm{z^-} \right|^2}
\end{equation}
and normalized residual energy
\begin{equation}
    \sigma_r = \frac{\left| \bm{v} \right|^2 - \left| \bm{b} \right|^2}{\left| \bm{v} \right|^2 + \left| \bm{b} \right|^2}.
\end{equation}
Here $\bm{v}$ and $\bm{b}$ are fluctuations of velocity and magnetic field (in Alfv\'en speed unit), and $\bm{z^\pm= \bm{u}\pm\bm{b}}$ are fluctuations of the two Els\"asser variables corresponding to outward and inward propagating Alfv\'en waves.
For perfectly Alfv\'enic fluctuations, we expect $\sigma_c = 1$ and $\sigma_r = 0$.
At 1 au, the fast solar wind typically shows high Alfv\'enicity, containing fluctuations with $\sigma_c > 0.5$ and $\sigma_r$ slightly negative, while the slow wind is usually non-Alfv\'enic with $\sigma_c$ close to zero and $\sigma_r$ close to -1, though slow Alfv\'enic solar wind is occasionally observed \citep{damicis2015origin}.
In addition, it has long been observed that the Alfv\'enicity of solar wind turbulence evolves as the solar wind propagates. Overall, both $\sigma_c$ and $\sigma_r$ drop with radial distance to the Sun, implying a decline of Alfv\'enicity \citep{tu1995mhd}. 

Figure \ref{fig:shi2021_fig08} shows $\sigma_c$ (left) and $\sigma_r$ (right) as functions of radial distance to the Sun (horizontal axis) and solar wind speed (vertical axis), calculated using PSP data from the first five encounters. 
$\sigma_c$ is high ($>0.5$) in general, indicating that the solar wind measured by PSP is mostly Alfv\'enic, except for the very slow populations $V_r < 250$ km/s, which are likely in the vicinity of heliospheric current sheet.
There is a clear decreasing trend of $\sigma_c$ with radial distance and a clear increasing trend of $\sigma_c$ with solar wind speed, implying that the Alfv\'enicity declines as the turbulence ages.
Except for the non-Alfv\'enic intervals when $\sigma_r$ is very small ($\sigma_r < -0.5$), $\sigma_r$ is in general slightly negative ($>-0.2$) and does not show significant variation with radial distance and wind speed.
Similar results are reported by other studies using PSP data \citep{chen2020evolution,sioulas2023magnetic,mcintyre2023properties}.
While theories and numerical simulations have shown that the negative residual energy is likely a natural result of the nonlinear evolution of MHD turbulence \citep{boldyrev2012residual,shi2023evolution}, in a homogeneous incompressible MHD turbulence system, $|\sigma_c|$ tends to gradually increase rather than decrease due to the ``dynamic alignment'' effect \citep{dobrowolny1980fully,dobrowolny1980properties}. 
Thus, to explain the decrease of $\sigma_c$ with radial distance, we need to go beyond the homogeneous incompressible turbulence system and consider additional effects, e.g. the spherical expansion of the solar wind that leads to reflection of Alfv\'en waves \citep{chandran2009,dong2014evolution,shi2023evolution}, large-scale velocity shears \citep{roberts1992velocity,shi2020propagation}, and parametric decay instability \citep{Tenerani2020}.

An important question is how do we relate the evolution of Alfv\'enicity of solar wind turbulence to the evolution of switchbacks.
\citet{Shi2022} show that the Alfv\'enicity of solar wind turbulence, manifested by the correlation between velocity and magnetic field fluctuations, is significantly lower in the quiescent intervals between switchback patches (see definition in Section \ref{sec:observation_occurrence}) than inside the switchback patches.
Therefore, a decrease of Alfv\'enicity, as shown by Figure \ref{fig:shi2021_fig08}, may indicate that the occurrence of switchback intervals declines as the solar wind propagates.

\section{Switchback Erosion Mechanisms}\label{sec:nonideal}

\subsection{Dispersive Effects at Switchback Boundaries}\label{sec:nonideal_boundaries}

In Sec.~\ref{sec:boundaries}, we discussed the way in which the inhomogeneity of the near-Sun solar wind may drive the formation of the observed sharp switchback boundaries, within the MHD framework. Once they form, specifically, if the boundaries steepen so much that their gradients become comparable to the ion inertial length $d_i = v_A/\Omega_i$ or the ion gyroradius $\rho_i = v_{th i}/\Omega_i$, MHD is no longer an applicable model, and two-fluid or kinetic effects become important to the evolution. \cite{mallet2023nonlinear} included two-fluid effects in an analysis of oblique large-amplitude Alfv\'en waves, showing that steady nonlinear solutions (e.g. solitons) can exist in which nonlinear steepening is balanced by dispersion: this result extends earlier work by \cite{kakutani1969weak}. In a homogeneous plasma, both dispersion and nonlinear steepening of Alfv\'en waves enter at small scales, and the solitons usually only exist at very large amplitudes: at smaller amplitudes and for general initial conditions, structures will tend to disperse over time.

\begin{figure}
    \centering
    \includegraphics[width=0.8\linewidth]{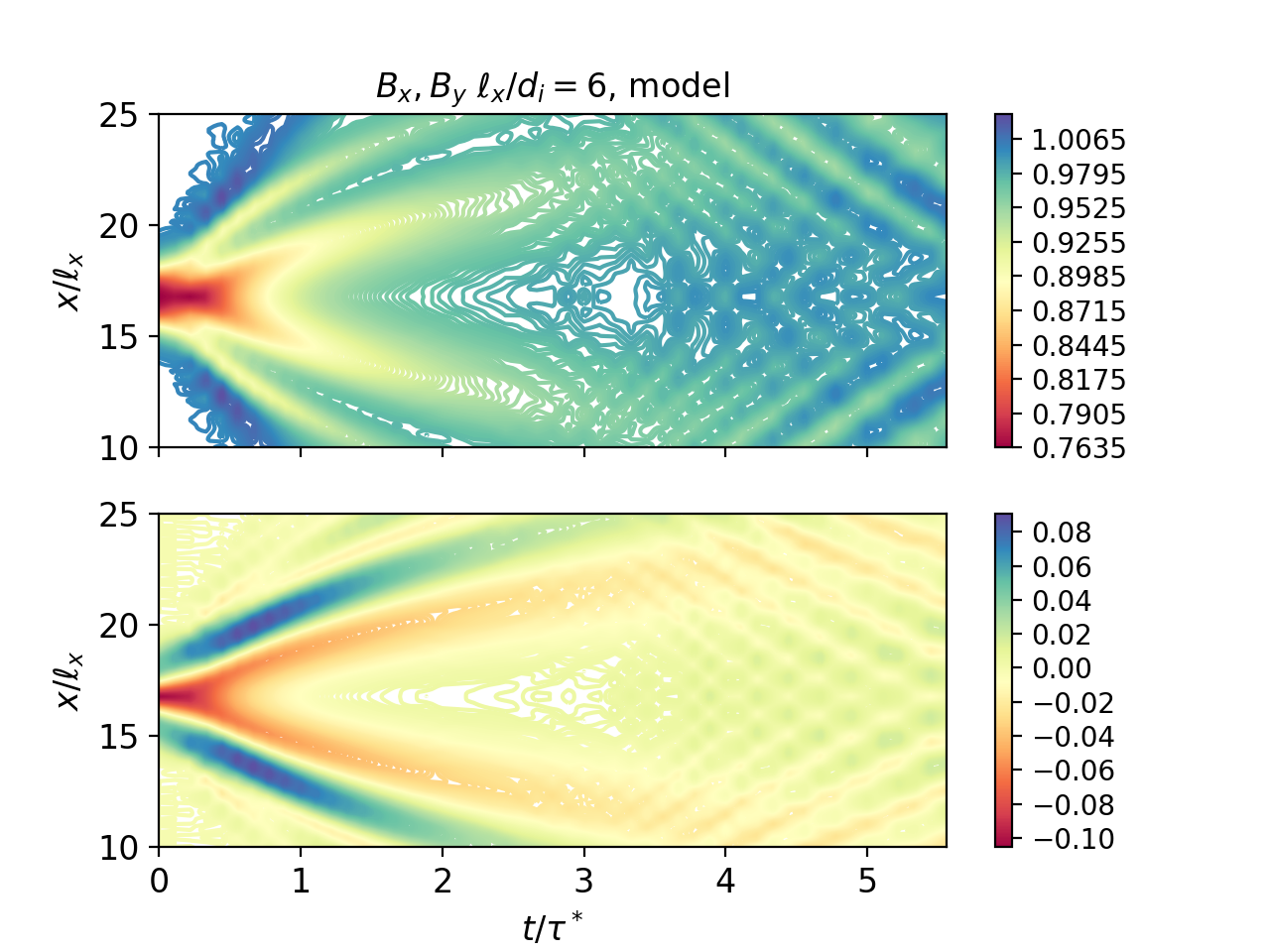}
    \caption{Contour plots of magnetic field components of an Alfv\'enic wave packet of length $\ell=6 d_i$ in a frame comoving at the Alfv\'en speed, showing the effect of dispersion in generating waves from the leading and trailing edges.  Figure reproduced with permission from \cite{tenerani2023dispersive}, copyright by AIP Publishing}
    \label{fig:tenerani-dispersive}
\end{figure}

In hybrid and Hall MHD numerical simulations, \citet{tenerani2023dispersive} show that, indeed, an initially coherent (small amplitude) wavepacket tends to disperse on the timescale $\tau^* \sim (l/v_A)(l/d_i)$, with a complex set of dynamics involving dispersion, coupling to compressible  modes and heating of the plasma. Figure \ref{fig:tenerani-dispersive} shows contour plots in the space-time plane of magnetic field components of an Alfv\'en wave packet, demonstrating that after a time of nearly $\tau^*$ the wave packet spreads due to the emission of waves ahead and behind the wave boundaries \citep{tenerani2023dispersive}. However, earlier simulations \citep{vasquez1996,vasquez1998} showed that in some situations, it is indeed possible for {combinations of monochromatic} Alfv\'en waves to steepen into steady discontinuities somewhat similar to the {soliton} solutions found by \cite{mallet2023nonlinear}, so it may be that the eventual steady-state depends sensitively on the initial conditions {of the wavepacket}. Moreover, all of the work incorporating small-scale steepening and dispersion of the waves does not take into account the growth in normalized amplitude and associated steepening caused by propagation in the inhomogeneous solar wind: it may be that the steady state in fact involves a balance between expansion-driven steepening \citep{Mallet2021,Squire2020,Squire2022} and small-scale dispersive effects.

Understanding the dynamics at switchback boundaries can help answer such questions. Besides work focused on determining the nature of discontinuities (see section \ref{sec:boundaries_observations}), there have been a number of studies dedicated to characterizing what type of waves or structures exist at switchbacks' boundaries that may contribute to their evolution.  Below we summarize relevant work on this topic by focusing on observations of small-scale waves at sharp boundaries, while reconnection (another mechanism that might contribute to the decay of switchbacks), is deferred to a later section.

\cite{Krasnoselskikh2020} first argued for the existence of surface currents and wave activity at the boundaries of switchbacks. 
They supposed a boundary layer formed at switchback boundaries due to strong density and ion plasma pressure gradients creating an effective electric field. {They showed that the minimum variance analysis on the boundary was consistent with a switchback embedded in a flux‑tube‑like magnetic structure, whose boundary carried a surface current flowing azimuthally (wrapping around the tube’s main axis)}. This would create a diamagnetic effect where the tube-aligned current would overcompensate for the quasi-perpendicularly aligned current, and produce a sheared magnetic field. Indeed, they found enhanced wave activity (signified by an increase in radial Poynting flux), at approximately the ion cyclotron frequency, at an oblique angle with respect to the boundary normal \cite{Krasnoselskikh2020, Larosa2021, choi_surface_2025}.  However, the reliability of the Solar Probe Cup (SPC) measurements \citep{case2020} that shaped this interpretation is not certain. Although quasi-thermal noise (QTN) was used for a more accurate density, SPC was used for a higher cadence density measurement along the boundaries, and the extent at which the particle properties fluctuate with respect to the instrumental effects is not certain. 
In spite of this, it became clear that adjacent switchback boundaries were home to currents and associated non-Alfv\'enic dynamics and wave emission, as investigated further in subsequent work. 

\begin{figure}
\includegraphics[width=0.99\textwidth]{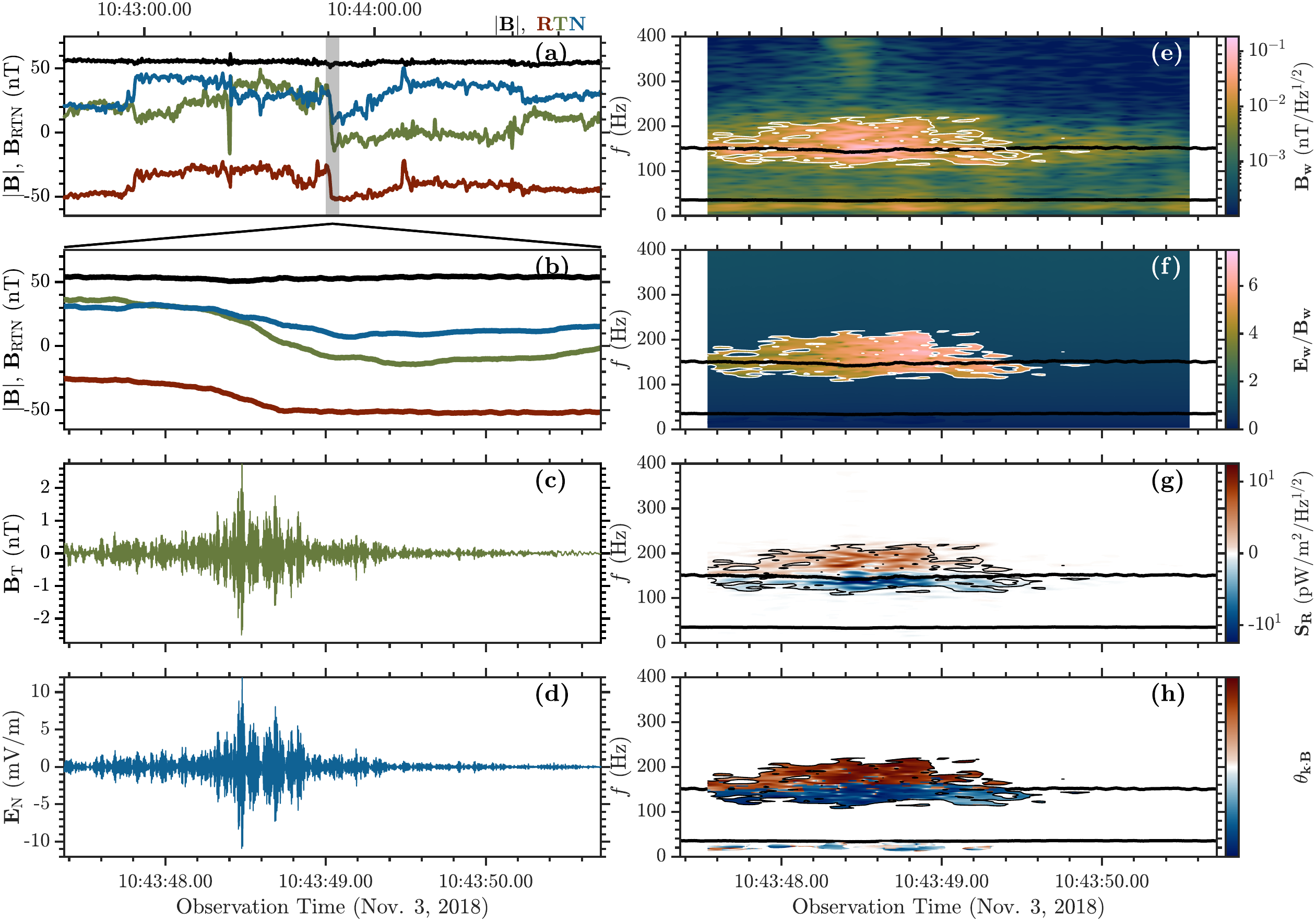}
\centering
    \caption{Whistler burst collocated with the switchback boundary detected at 10:34:18.22 UTC on November 3, 2018: (a) Dynamics of the magnetic field in $\mathrm{RTN}$ coordinates from the MAG over a short time interval around the burst; (b) The zoomed shaded region of (a) shows the magnetic field only during the burst; (c) SCM burst waveform of the magnetic field $\mathrm{T}$-component, $\delta B_\mathrm{T}$; (d) The burst EFI electric field $\mathrm{N}$-component, $\delta E_\mathrm{N}$; (e) The magnetic field dynamic spectrum; (f) The ratio of electric spectrogram to magnetic spectrogram with the background colouring representins the expected ${E_w/B_w}$ ratio for parallel propagating whistler waves; (g) The radial component of the Poynting flux; (h) Wave normal angle with respect to the background magnetic field, $\theta_{\mathbf{k\cdot B}}$, ranging from $0^\circ$ to $180^\circ$ and indicating parallel and anti-parallel propagation, respectively. The lower and upper solid black curves in (e)-(h) indicate $f_{\mathrm{lh}}$ and $0.1f_{\mathrm{ce}}$, respectively. Figure reproduced with permission from \cite{karbashewski2023}, copyright by AAS}
    \label{fig:wistlers_at_SB_boundaries}
\end{figure}

\cite{Farrell2020} and \cite{Rasca2021} studied the structure of the boundaries of the sharpest switchbacks corresponding to full magnetic field reversals, finding evidence of magnetic field drop-outs (magnetic holes) a few seconds long. Like \cite{Krasnoselskikh2020}, \cite{Farrell2020} also argued that the change in particle pressure across the switchback boundary could lead to a diamagnetic current leading to such drops in the magnetic field strength. 
However, no significant wave activity was found associated with the sharpest boundaries. A later analysis of a larger dataset of switchbacks showed that there exist some with degraded boundaries with associated wave activity \citep{farrell2021switchback}. These observations point to some aging process of switchbacks accompanied by wave activity.  While \cite{farrell2021switchback} specifically identified ULF (ultra-low-frequency) waves (lasting a few seconds), other work found evidence of higher frequency wave activity. These include waves at nearly the plasma frequency, often associated with electron beams \citep{Rasca2022}, whistler waves driven presumably by the perturbation of the suprathermal electron population distribution function \citep{froment2023,karbashewski2023,agapitov2020, colomban_polarization_2025}, {and kinetic Alfv\'en waves \citep{malaspina2022inhomogeneous}}. Figure \ref{fig:wistlers_at_SB_boundaries} shows counter-propagating whistler bursts located at the switchback boundary and indicating crossing the wave generation region \citep{karbashewski2023, choi_whistler_2024, vo_enhanced_2024}. 

Although more statistical and theoretical work is required to fully understand the origin of such waves and their relation to switchback evolution, these results indicate that kinetic scale behavior at the boundaries of switchbacks may be contributing to their decay and overall stability.   

\subsection{Magnetic Reconnection}\label{sec:reconnection}
\subsubsection{Direct Observations of Reconnection and Implications for Switchback Evolution}
\begin{figure}
\includegraphics[width=\textwidth]{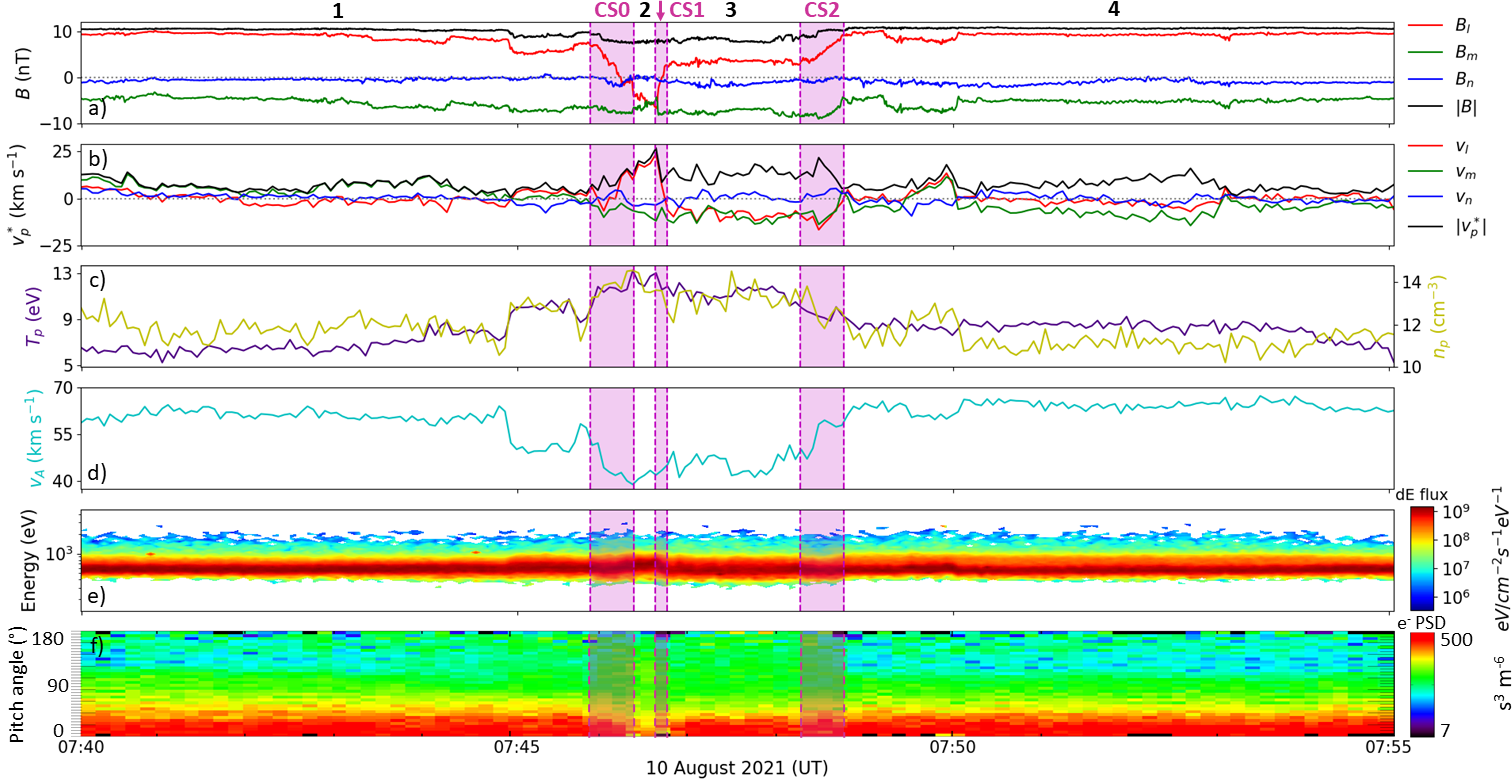}
\centering
    \caption{Example of a reconnecting switchback observed by SO in the MVA-frame. a) Magnetic field vector with the magnetic field strength in black. b) Proton bulk velocity with the proton bulk speed in black. The average proton bulk velocity $<\mathbf{v_p}>$ over this interval has been removed. In both panels, the l-component is in red, the m-component is in green, and the n-component is in blue. c) Proton temperature (left scale, purple) and number density (right scale, gold). d) Alfvén speed. e) 1D proton energy spectrogram. f) Electron strahl PAD for energies $>$70 eV. The shaded regions show the locations of the current sheets at the switchback boundaries. Figure reproduced with permission from \citet{Suen2023}, copyright by ESO}
    \label{Fig3.5A}
\end{figure}

One important process that may contribute to the evolution of the sharp boundaries of switchbacks is magnetic reconnection: a fundamental energy conversion process in plasmas, where particles are heated and accelerated by the {magnetic energy} released by changes in the magnetic field topology. Using PSP observations, \citet{AkhavanTafti2021} show that the proton beta and magnetic shear angle at switchback boundaries theoretically favour reconnection \citep{Swisdak2003}. However, observations of reconnecting switchbacks from PSP \citep{Froment2021} and SO \citep{Fedorov2021, Suen2023} are limited, showing that switchback boundary reconnection is a rare occurrence.

Figure \ref{Fig3.5A} shows an example of a reconnecting switchback observed by SO on 10 August 2021, when the spacecraft was 0.72 au from the Sun. Here, one can see the heliospheric magnetic field (HMF) polarity reversal and electron strahl pitch angle distribution signature commonly associated with switchbacks. The leading edge boundary of the switchback is non-reconnecting and consists of a single current sheet, labeled CS0. The trailing edge boundary has undergone reconnection and developed into a bifurcated reconnection current sheet \citep{Gosling2005}, denoted as CS1 and CS2. The reconnection outflow is confined to the region between these two current sheets. Although the case presented in this figure shows reconnection at the trailing edge boundary, there does not appear to be a preference for whether reconnection occurs at the leading or trailing edge boundary of the switchback. The velocity enhancement inside the switchback is $35\%$ of the local Alfv\'en speed. This is smaller than in typical non-reconnecting switchbacks, and these sub-Alfv\'enic velocity enhancements are a common characteristic of all reconnecting switchbacks \citep{Froment2021, Suen2023}. A more detailed discussion of the properties of reconnecting switchbacks may be found in \citet{paper3}.

\begin{figure}
\includegraphics[width=0.6\textwidth]{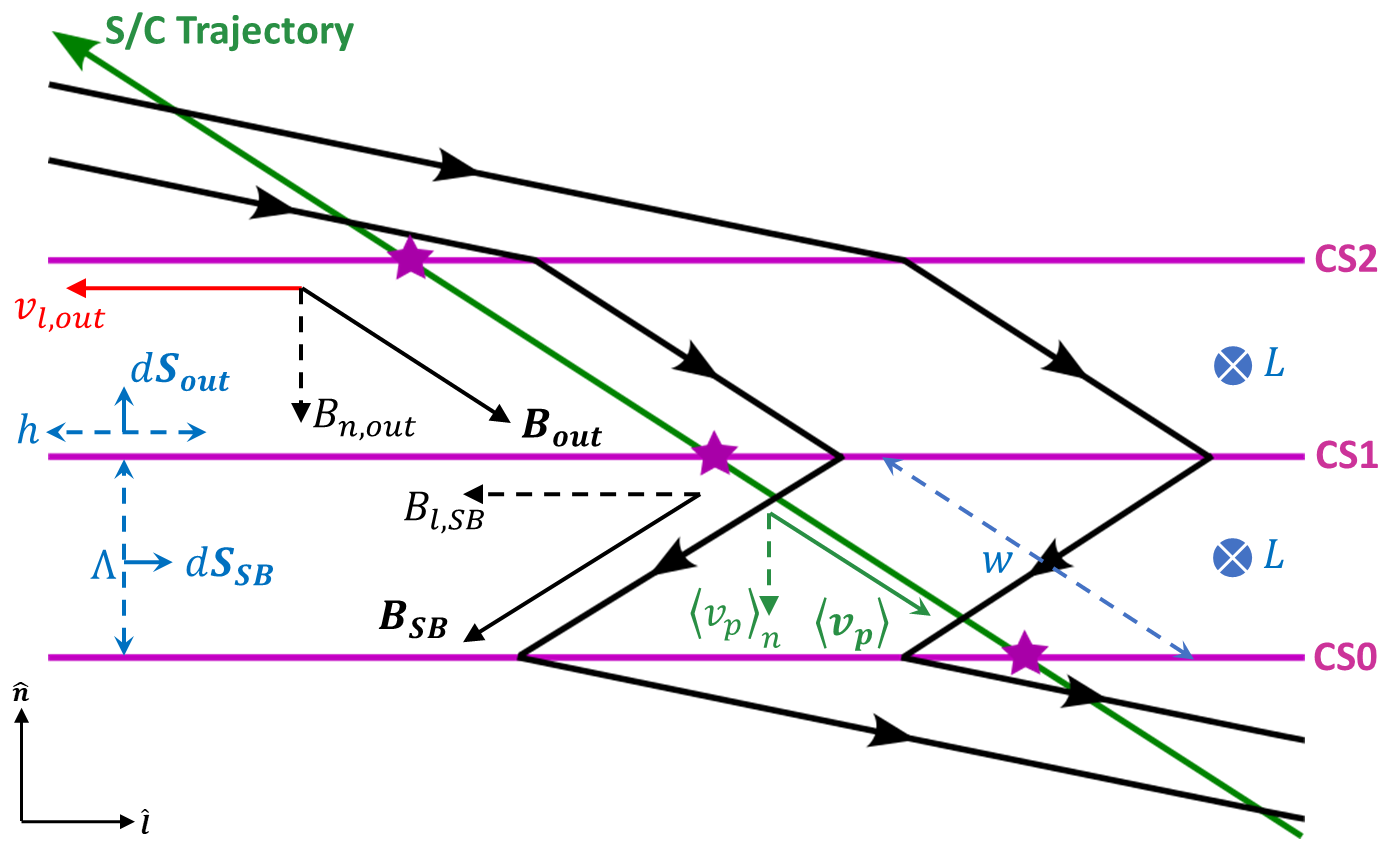}
\centering
    \caption{Schematic diagram of the geometry of a reconnecting switchback with parameters used to estimate $\tau$ defined. The black arrows show the magnetic field configuration of the switchback, and the green arrow shows the spacecraft trajectory through the structure. The purple lines show the current sheets CS0, CS1, and CS2 at the switchback boundaries, and the purple stars mark where the spacecraft crosses these current sheets. The reconnection outflow is shown by the red arrow. Reproduced with permission from \citet{Suen2023}, copyright by ESO}
    \label{Fig3.5B}
\end{figure}

Reconnection may accelerate the erosion of magnetic switchbacks by removing magnetic flux from the interior of the switchback where the magnetic field polarity is reversed. Figure \ref{Fig3.5B} is a simplified diagram of the magnetic field geometry of the switchback shown in Figure \ref{Fig3.5A}. From the definition of magnetic flux $\phi = \int\mathbf{B}\cdot d\mathbf{S}$ through a surface comprised of infinitesimal surface elements $d\mathbf{S}$, the magnetic flux remaining in the polarity-reversed section of the switchback, $\phi_{SB}$ is:
\begin{equation}
    \phi_{SB} = B_{l,SB}\Lambda L,
    \label{Eq3.5A}
\end{equation}
where $B_{l,SB}$ is the component of the magnetic field inside the switchback parallel to the current sheets; $L$ is the out-of-plane extent of the switchback; and $\Lambda$ is the perpendicular distance between CS0 and CS1, used to estimate the width of the switchback. Similarly, the rate of flux transport by the reconnection outflow $\dot{\phi}_{out}$ is:
\begin{equation}
    \dot{\phi}_{out} = B_{n,out}v_{l,out}L,
    \label{Eq3.5B}
\end{equation}
where $B_{n,out}$ is the component of the magnetic field in the outflow region normal to the current sheets and $v_{l,out}$ is the current sheet-aligned component of the reconnection outflow.

\citet{Suen2023} performed a timing analysis of this process to evaluate the viability of magnetic reconnection as an erosion mechanism for magnetic switchbacks. The switchback erosion timescale, $\tau$, is defined as the ratio between $\phi_{SB}$ and $\dot{\phi}_{out}$:
\begin{equation}
        \tau = \frac{\phi_{SB}}{\dot{\phi}_{out}} = \frac{B_{l,SB}\Lambda}{B_{n,out}v_{l,out}}.
        \label{Eq3.5C}
\end{equation}
Estimates of $\tau$ for three examples of reconnecting switchbacks observed by SO at a heliocentric distance of 0.6-0.7 au show the erosion timescales range from 40 minutes to $~2000$ minutes (Table \ref{Tab3.5}. For a typical slow solar wind stream with an average speed $|\langle\mathbf{v_p}\rangle|$ of 300-400 km s$^{-1}$, the distance $D \simeq |\langle\mathbf{v_p}\rangle|\tau$ traversed by the switchback until its complete erosion is between 0.005 to 0.4 au.
{We note that theoretically the local Alfv\'en speed $V_A$ needs to be considered in the estimate of $D$ since switchbacks are mostly outward propagating Alfv\'enic structures. However, the events analyzed by \citet{Suen2023} are at distances larger than 0.6au where $V_A$ is much smaller than the local solar wind speed. Thus, the inclusion of $V_A$ will increase $D$ only slightly (on the scale of 10\%).}

\begin{table}[h]
    \caption{Estimates for switchback erosion timescale $\tau$. Table reproduced with permission from \citet{Suen2023}, copyright by AAS, where the exact timings/distances for the events 1-3 may be found.}\label{tab1}%
    \begin{tabular}{@{}lllllllll@{}}
    \toprule
        Event & $\Lambda$ (km) & $B_{l,SB}$ (nT) & $B_{n,out}$ (nT) & \makecell{$v_{l,out}$ \\(km s$^{-1}$)} & \makecell{$|\langle\mathbf{v_p}\rangle|$ \\(km s$^{-1}$)} &$\tau$ (min) & $D$ (au)\\
        \midrule
        1 & 3570 & -4.8 & -1.0 & -7.2 & 322 & 40 & 0.005  \\
        2 & 10100 & -9.7 & 0.5 & -28.3 & 439 & 126 & 0.02  \\
        3 & 31700 & -7.5 & -0.1 & 15.4  & 443 & 2005 & 0.4  \\
    \botrule
    \end{tabular}
    \label{Tab3.5}
\end{table}

The estimates for $\tau$ for all three switchbacks are shorter than the characteristic timescales for solar wind expansion, showing that when it occurs, reconnection is a fast and efficient mechanism for switchback erosion, {provided that it remains efficient until the switchback is completely destroyed.} Furthermore, the estimated $D$ also indicates that full erosion of the switchbacks occurs at or before 1 au. The short timescales over which this process occurs could explain why so few examples of reconnecting switchbacks have been reported as the window of opportunity to observe them is brief. However, the exact time and location of reconnection onset are unknown -- the use of single-spacecraft measurements in this study limits knowledge of the time history of switchbacks.

If one assumes that the reconnection rate remains constant and that reconnection onset occurs closer to the Sun than SO (roughly $\sim0.1$ au, comparable to Parker Solar Probe Encounter 1), one can show that the extrapolated switchback width $\Lambda$ at these distances would be significantly greater than what has actually been observed. This result has two implications on the origins and evolution of switchbacks during transport in the solar wind. If the reconnecting switchbacks are formed in the solar atmosphere, they must have remained stable out to $\sim0.6$ au before reconnection is triggered at the boundaries during transport in the solar wind, resulting in the rapid erosion of the structure. The suppression of reconnection at the switchback boundaries by mechanisms such as shear flows \citep{Owen1987, Chen1990, Cassak2011} and diamagnetic drifts \citep{Swisdak2003, AkhavanTafti2021} may play a role in maintaining the stability of the structure with respect to reconnection onset during transport. 

Alternatively, the reconnecting switchbacks may have formed \emph{in situ} in the solar wind and thus have been extant for a shorter time before reconnection onset. The implication of \emph{in situ} generation of switchbacks from this scenario supports results from \citet{Macneil2020} and \citet{Pecora2022} that show the occurrence rate of switchbacks increases with heliocentric distance, rather than decreasing as one would expect if switchback formation occurred solely in the corona. The latest PSP encounters have also shown a drop in switchback observations near the Sun, especially in the solar corona. 

In conclusion, when it occurs, reconnection is an efficient mechanism for removing individual switchbacks from the solar wind. However, its overall effectiveness as a switchback erosion mechanism in the heliosphere depends on how frequently switchback boundary reconnection occurs. The scarcity of reported switchback reconnection events suggests that this phenomenon is rare and that reconnection is suppressed at switchback boundaries. Strong flow shears across current sheets are known mechanisms for suppressing reconnection \citep{Chen1990,Cassak2011,Mallet2025} and may explain why reconnecting switchbacks are almost always associated with sub-Alfv\'enic velocity enhancements. 

\subsubsection{Switchback Evolution as Flux-rope Merging and Reconnection}

One model for switchback generation is that they are flux ropes formed in the low corona via interchange reconnection, and subsequently injected into the solar wind \citep{Drake2021}. Such flux ropes are generated at small spatial scales as part of reconnection \citep{biskamp1986magnetic,drake2006electron,cassak2009, choi_switchbacks_2025}. The magnetic island (wrapped magnetic field) structure of the flux ropes, close to unity aspect ratio, and low Alfv\'enicity is very different from the known properties of switchbacks in the solar wind \citep{Drake2021}, leading to a challenge for this model of switchback formation.

\begin{figure}
\includegraphics[width=0.99\textwidth]{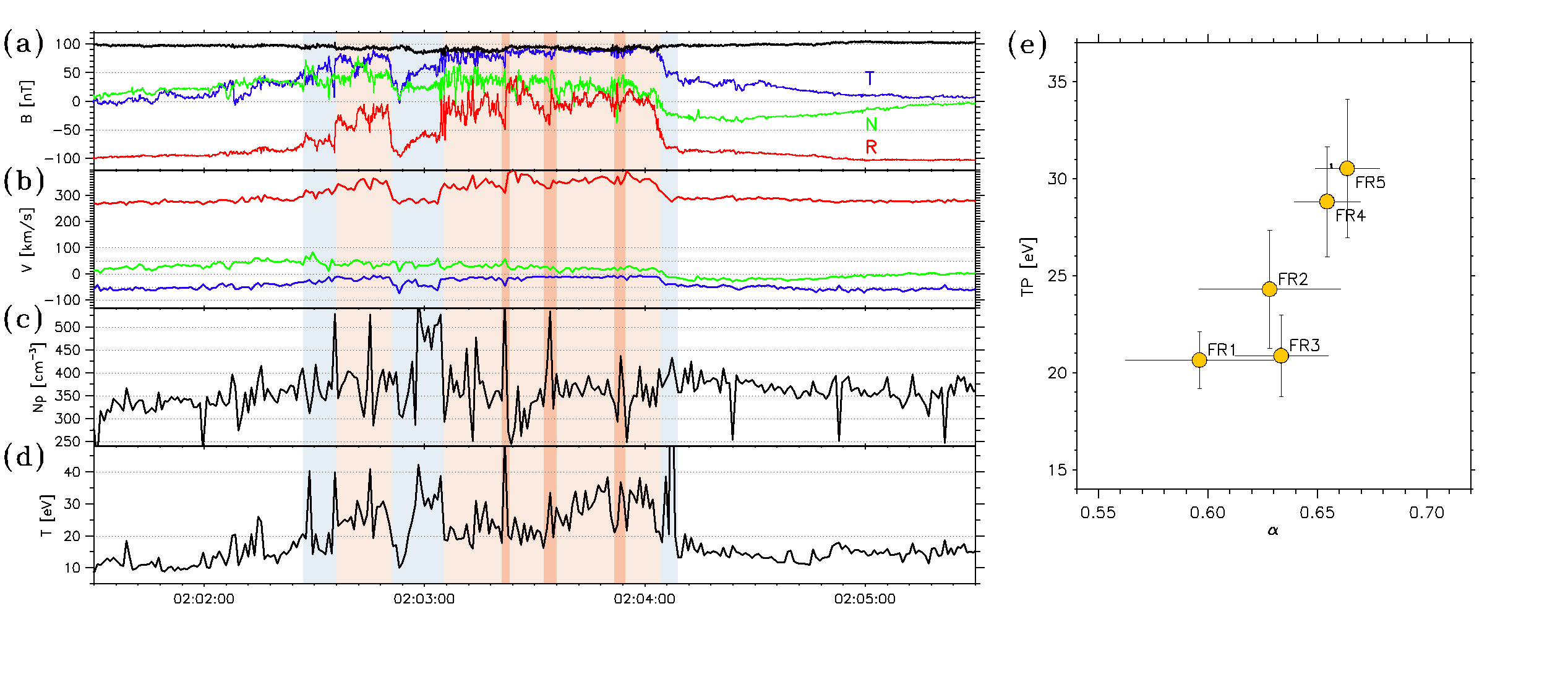}
\centering
    \caption{Two switchbacks (highlighted by light-red) recorded on November 6, 2018 by Parker Solar Probes. The panels from top to bottom present the magnetic field in the RTN coordinate system - (a) and the proton bulk velocity - (b); the proton density - (c) and the parallel proton temperature - (d). The deep red regions highlight boundaries between distinct regions of the second switchback. (e) The dependence of the proton temperature on the Alfv\'enicity for the structure components (individual flux ropes) composing the switchback. Figure reproduced with permission from \citet{Agapitov2022}, copyright by AAS}
    \label{fig:issi_psp_mering}
\end{figure}

\citet{Agapitov2022} propose that, while flux ropes form with an aspect ratio of order unity and comparable axial and transverse magnetic field, while propagating outward in the solar wind, they interact with each other through merging of similar flux ropes generated from the same source. During the merging process, reconnection erodes the strength of the wrapping magnetic field and heats the plasma inside the structure \citep{drake2006electron,zhou2014plasma,zhou2017coalescence}, an energetically favorable process.
This controls the elongation of flux ropes: a weaker wrapping magnetic field allows the ambient solar wind magnetic field to squash and elongate the flux ropes (and therefore switchbacks). Thus, switchbacks become increasingly elongated along the solar wind magnetic field with radial distance from the Sun, potentially explaining the observed geometry \citep{Laker2021}. 
A weak wrapping field and strong axial field also results in a sharp boundary rotation of the magnetic field, as observed.
Moreover, merging also has the effect of increasing the axial plasma flow speed, leading to increased Alfv\'enicity. This, therefore, would cause flux ropes to evolve towards the switchback-like structures \citep{Drake2021}. Other features of merged flux ropes seen in simulations, including localized current layers and localized density and temperature enhancements, are often seen in the switchback structures observed in the solar wind by PSP. Some observational evidence for the early stages of a switchback interaction that may lead to merging is presented in Figure \ref{fig:issi_psp_mering}. Two switchbacks (highlighted in red) are approaching each other and driving a density enhancement between them (highlighted in blue). 

\begin{figure}
 \includegraphics[width=0.99\textwidth]{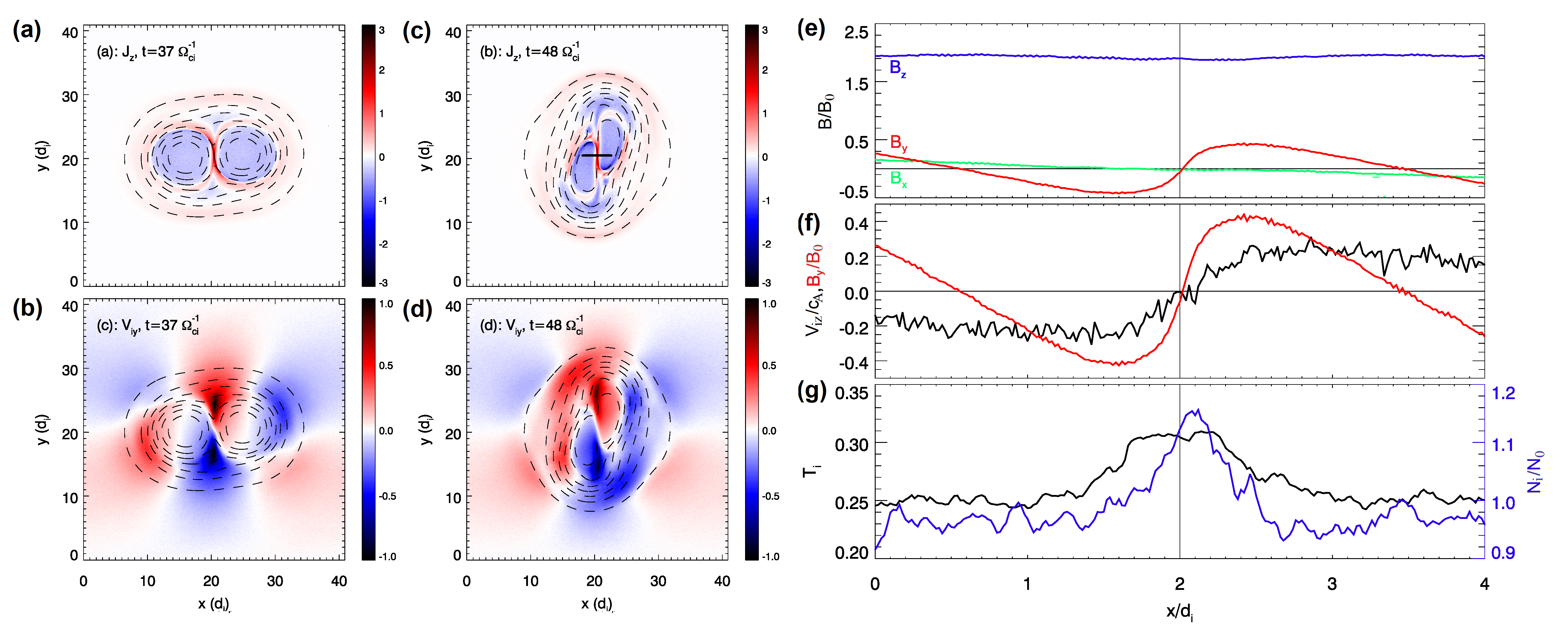}
 \centering
     \caption{Results of a simulation of flux rope merger with initial Alfv\'enic flow: (a) Out-of-plane current and magnetic field lines during merging and (c) after merging ends; (b) and (d) Vertical plasma flows for the corresponding times in (a) and (c). The data along the black line in panel (c) is shown in panels (e-g): (e) The magnetic field; (f) The reversing magnetic field component (the red curve) and the corresponding component of the plasma flow velocity (the blue curve); (g) The plasma density (the blue curve) and temperature (the black curve). Figure reproduced with permission from \citet{Agapitov2022}, copyright by AAS}
     \label{fig:issi_psp_mering_marc}
 \end{figure}

When the Alfvénicity approaches unity, merging becomes energetically unfavorable: the saturation of flux rope merging is a consequence of energy transfer from the reconnected magnetic field into the plasma flow and thermal energy. If the flux rope Alfvénicity becomes significant during a merger, merging can be arrested before it is complete (see simulations presented in Figure.~\ref{fig:issi_psp_mering_marc}), leading to remnant current sheets with magnetic and velocity characteristics consistent with PSP observations of multi-flux-rope structure in switchbacks (Figure \ref{fig:issi_psp_mering}). 

This model suggests that many of the observed features of switchbacks can be explained by multiple merging events between nearby flux ropes. However, significant questions remain: the increase in Alfv\'enicity due to mergers must be reconciled with the overall decrease in Alfv\'enicity with radial distance (see Sec.~\ref{sec:alfvenicity}). One potential signature of flux ropes merging controlling the evolution of switchbacks would be a relationship between plasma temperature and Alfv\'enicity inside a switchback, since both increase during merging.

\subsection{Parametric Decay of Switchbacks}\label{sec:pdi}


It is well known that, although a circularly-polarized Alfv\'en wave with uniform density and pressure is an exact solution to the MHD equation set, it is susceptible to the parametric decay instability \citep[PDI, e.g.][]{derby1978modulational, goldstein1978instability, hollweg1994beat, del2001parametric}, which makes the initial Alfv\'en wave decay into a forward propagating sound wave and a backward propagating Alfv\'en wave. 
As the observed switchbacks are mostly spherically polarized Alfv\'en waves, PDI is potentially an important mechanism in dissipating the switchbacks.

\begin{figure}[htb!]
    \centering
    \includegraphics[width=\textwidth]{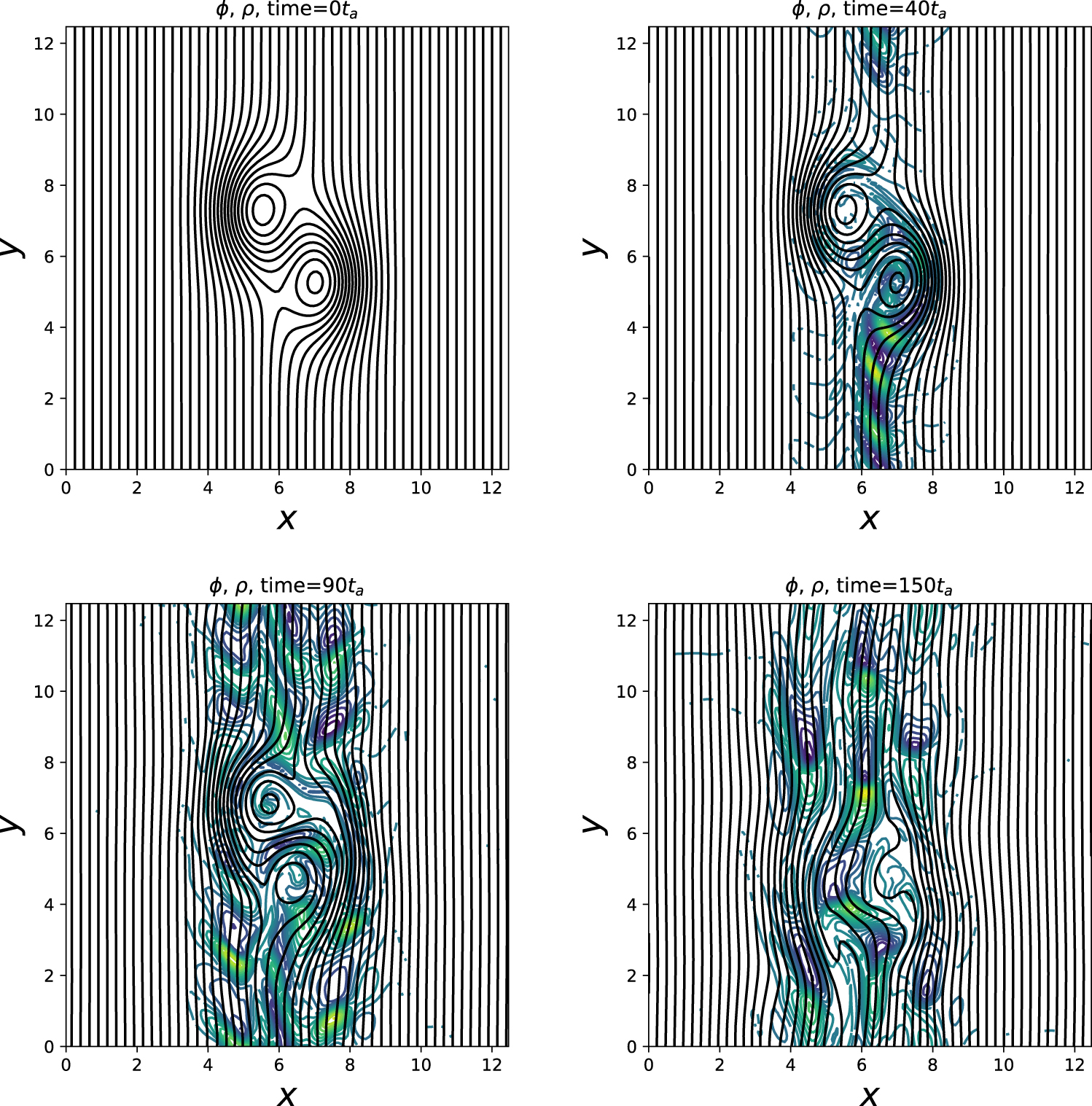}
    \caption{Evolution of a 2D switchback in a MHD simulation. The black lines are magnetic field lines, and the colored lines are contours of the density perturbation. Figure reproduced with permission from \citet{Tenerani2020}, copyright by AAS}
    \label{fig:tenerani2020_fig03}
\end{figure}

\citet{Tenerani2020} and \citet{marriott2024parametric}, by 2D and 3D MHD simulations, analyzed the stability problem of a 2D Alfv\'en wave packet, as shown in the top left panel of Figure \ref{fig:tenerani2020_fig03}. The figure displays the time evolution of magnetic field lines and the contours (colored lines) of density perturbation. 
We can clearly see the growth of the density fluctuation as the simulation runs, indicative of the parametric decay instability.
Nonetheless, the switchback persists for a quite a long time (at least 100$t_a$ with $t_a$ being the Alfv\'en crossing time) and is destroyed much later than the saturation of the PDI, which happens at $t \sim50 t_a$. 
Note that the simulation shown by Figure \ref{fig:tenerani2020_fig03} has $\beta = 0.1$ that is low and more applicable to the near-Sun region.
Thus, \citet{Tenerani2020} concluded that it is possible that the observed switchbacks are generated in the lower corona \citep{Telloni2022} and survive until the location of PSP. 

In a later work by \citet{shi2024analytic}, 3D MHD simulations of a total switchback, i.e. $\Delta B_r /B_0 = 2$, are conducted to investigate the effect of the switchback's geometry and plasma beta on its stability.
{They verify that the switchback is stable if incompressible MHD is utilized, suggesting the growth of the PDI is a primary mechanism capable of destabilizing the structure within the MHD framework.}
\citet{shi2024analytic} conclude that the growth rate of PDI is smaller if the structure is longer in the background magnetic field direction or is more 2D like. 
It is also found that the growth rate of PDI is not a monotonically decreasing function of $\beta$, in contrast to the prediction by linear theory of a monochromatic circularly polarized Alfv\'en wave.
\cite{shi2024analytic} show that the growth rate of PDI is larger for $\beta \ll1$ and $\beta > 1$ than $\beta \lesssim 1$. 
In addition, the pattern of the density fluctuation is quite different depending on $\beta$. For $\beta > 1$, the growing density fluctuation is mainly concentrated inside the switchback region, while for $\beta < 1$, the density fluctuation is distributed along the background field direction.
For $\beta > 1$, the folded magnetic field lines straighten out soon after the PDI saturates, while for $\beta < 1$, the field lines maintain folded for a much longer time after the saturation of PDI, similar to the result of \citet{Tenerani2020}.

{We need to point out that, neither of the two analytic models mentioned above can capture all the observational characteristics of the switchbacks. One major problem is that the analytic models cannot control whether a magnetic field line connects to infinity or is closed locally, and thus both the models contain structures like flux ropes or magnetic islands. In contrast, it is believed that the realistic switchbacks primarily consist of open field lines with one end connected to the Sun and one end open to infinity as indicated by the unidirectional strahl electrons. The other problem is that, both of the two models cannot fully confine the switchbacks within a local region. Even in the 3D model developed by \citet{shi2024analytic} where the switchback seems to be have a finite extent in all directions, the background magnetic field is twisted along the azimuthal direction out of the switchback region. \citet{shi2024analytic} has proven that, for a spherically polarized magnetic field, it is impossible to have a localized switchback without changing the background magnetic field outside the region due to the $\nabla \cdot \bm{B}=0$ condition. This suggests that $|B|$ may not be perfectly uniform throughout a realistic switchback. Further investigation on this topic is necessary.}

We note that, as the size of the switchback approaches the ion inertial scale, the Alfv\'enic status is no longer an exact solution to the system, and the kinetic effect facilitates the disruption of the switchbacks \citep{tenerani2023dispersive}. 
In this case, the proton internal energy increases because of contributions from both the compressive process and phase space mixing due to dispersion of the wave modes. 
As a final remark, all the previous works rely on local simulations by assuming homogeneous background plasma. However, the spherical expansion effect of the solar wind is a non-negligible effect because it de-correlates the magnetic field and velocity \citep{grappin1996waves}, reducing the Alfv\'enicity of the structure. 
Thus, it would be important to investigate how the expansion effect modifies the evolution of switchbacks in the future.

\subsection{Interaction with Interplanetary Shocks }\label{sec:IP_shocks}

The collisionless nature of the solar wind raises open questions as to how the turbulent cascade of energy is transferred between disparate scales and eventually converted into plasma heat. Owing to their ability to accelerate particles to high energies \citep{Reames1999}, interplanetary (IP) shocks represent one possible mechanism for the dissipation of energy, exchanging plasma flow energy upstream to magnetic energy downstream. IP shocks are mostly driven by Coronal Mass Ejections \citep[CME, ][]{Sheeley1985} and Stream Interaction Regions \citep{Richardson2018}, both of which have been shown to highly influence geoeffectiveness. Unravelling the highly complex nature of how shocks interact with their surrounding environment (and vice versa) is contingent upon assessing the universality of energy partition and exchange processes across the shock. Which turbulence properties are invariant upstream versus downstream of a shock? In the high beta case, for example, \citet{Zank2021} found that the inertial range spectral index and overall shape of the spectrum were largely unchanged across the shock, while the amplitude of the fluctuations was significantly higher downstream compared to upstream. However, the precise nature of how turbulence, or any general structure or wave interacts with a collisionless shock is largely unknown \citep{Guo2021}. 

\citet{trotta2024properties} addresses the question of how switchbacks (one-sided Alfv\'enic deflections of ``moderate amplitude") are processed through a strong CME-driven fast-forward shock seen with PSP (0.07 au) on September 5th, 2022. It was found that the Alfv\'enic nature of switchbacks were preserved across the shock, yet with increased small-scale compressive fluctuations downstream. Although the degree of high cross helicity was preserved, the polarization arc of the switchbacks changed from the R-T plane upstream to both R-T and R-N plane downstream, suggesting that the switchbacks were compressed along the normal and stretched along perpendicular direction. Shocks may thus impact switchback lifetime by introducing dispersive effects, eventually eroding their boundaries. 

\par The spectral information transmitted across shock was found to be consistent with prior observations and expectations \citep{Zank2021}. In addition, the inertial range was recovered in the downstream, as opposed to observations of the bow shock ``resetting" the turbulence cascade \citep{sahraoui2020magnetohydrodynamic}, questioning the universality of the existence of the inertial range across plasma regimes. The downstream spectrum also shows a fluctuation enhancement of 4 times the upstream, and a flattening around the ion cyclotron scale. Although the disappearance of the ion cyclotron wave (ICW) signatures downstream may only be due to high $\theta_{BV}$ obscuring the measurement \citep{bowen2020a}, \citet{trotta2024properties} points out that there is no proton beam observed to participate in the generation of ICWs, as expected in a more turbulent downstream environment \citep{valentini2010two}. However, the protons went out of the SPAN-I field-of-view in the downstream region, so the existence of the proton beam is unknown.

\par \cite{trotta2024properties} cautions that at PSP, the fluctuation level downstream may be overestimated due to the observed extreme velocity jump across the shock. The broad distribution of switchback properties \citep{Laker2021} and single-point measurement limitation make it difficult to assess how the same plasma parcel is processed across the shock. In addition, the spacecraft frame velocity was much higher downstream, so the switchbacks may appear shorter, and the observed increased switchback deflection amplitude may be a consequence of large {$|\boldsymbol{B}|$}. The high level of shock variability and other assumptions pose additional caveats that may lead to erroneous shock characterization.

\par Finally, \cite{trotta2024properties} also describes the radial evolution of the CME-driven shock seen with PSP (0.07 au) and with Solar Orbiter (0.7 au). Although the shock crossing orientation changed from quasi-perpendicular to quasi-parallel, a higher level of structure and variability was found downstream from the shock at Solar Orbiter, implying an increasing level of complexity with increasing radial distance. Upstream from the shock at Solar Orbiter, no switchbacks were present, but rare ``shocklets" were observed instead \citep{Trotta2023b}.

\section{Impacts of Switchbacks}\label{sec:impacts}

\subsection{Turbulence }\label{sec:turbulence}


Before talking about the impacts of magnetic switchbacks on the solar wind turbulence, one important question to discuss is \textit{whether we can distinguish switchbacks and turbulence}, i.e. whether switchbacks are part of the turbulence or standalone structures.

\begin{figure}
    \centering
    \includegraphics[width=0.5\hsize]{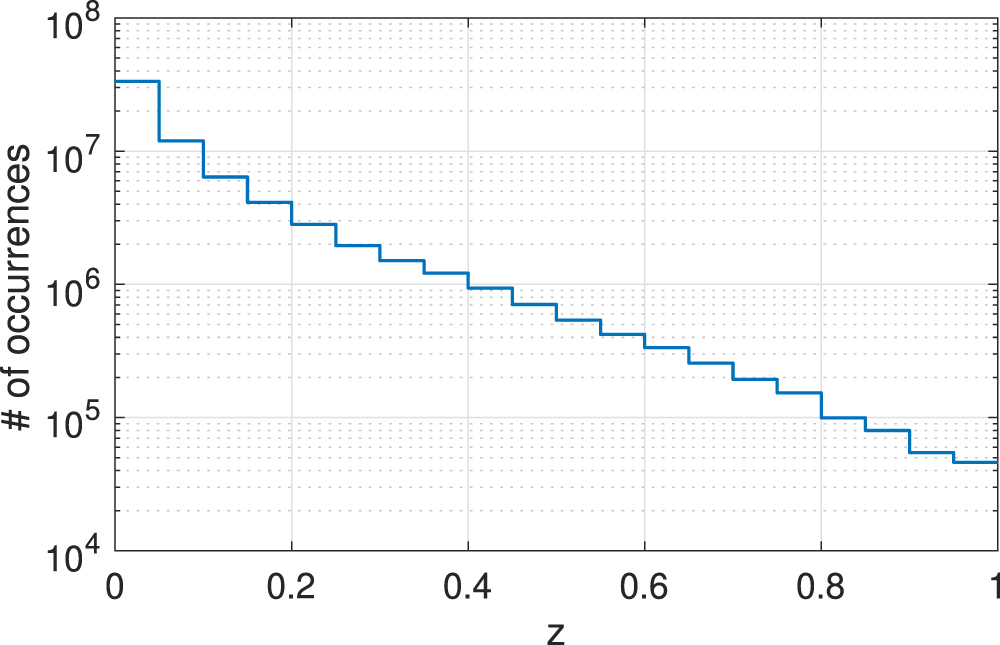}
    \caption{Histogram of the values of z for the 10-day time interval from Nov 01 to Nov 10, 2018. Figure reproduced with permission from \citep{DudokdeWit2020}, copyright by AAS}
    \label{fig:dudokdewit2020_fig04}
\end{figure}

PSP observations indicate the possibility that switchbacks are simply large-amplitude turbulence.
A major piece of supporting evidence is the continuous distribution of the $z$-value,
\begin{equation}
    z = \arccos\left(\frac{\bm{B}\cdot \bm{B_0}}{|\bm{B}|| \bm{B_0}|}\right),
\end{equation}
which quantifies the angle between the instantaneous magnetic field vector and the background magnetic field vector \citep{DudokdeWit2020}, as shown by Figure \ref{fig:dudokdewit2020_fig04}.
If switchbacks are standalone structures instead of part of the turbulence, we expect to see a bi-modal distribution of the $z$-value.
In addition, numerical simulations \citep{Squire2020,Shoda2021} confirm that WKB evolution of Alfv\'en waves in the expanding solar wind is able to generate switchbacks. The readers are encouraged to refer to \citet{paper5} for more information on the generation mechanisms of switchbacks.

\begin{figure}
\includegraphics[width=0.6\textwidth]{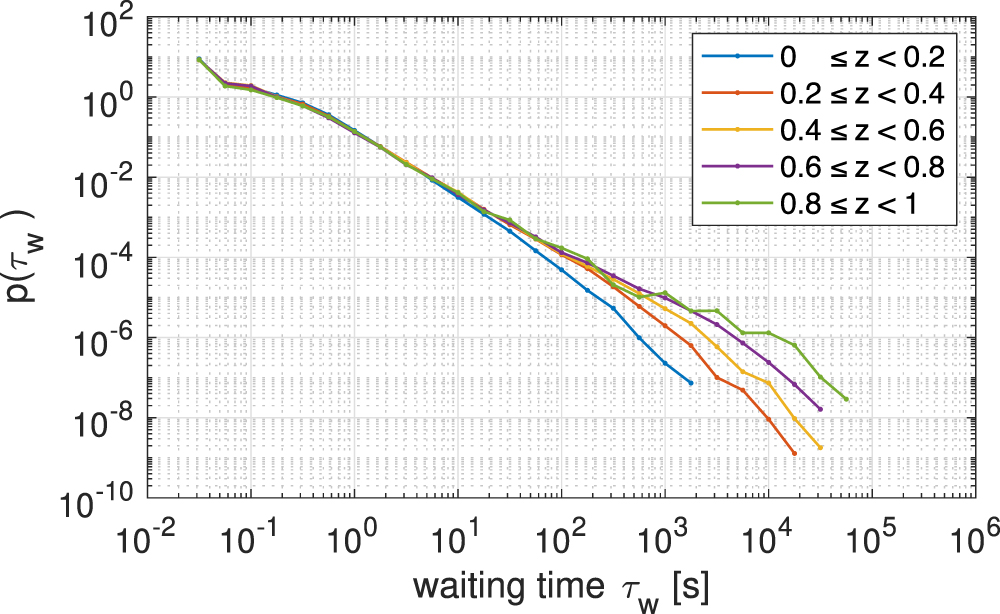}
\centering
    \caption{ Waiting time distribution of normalized deflection $z$ for different thresholds. Figure reproduced with permission from \cite{DudokdeWit2020}, copyright by AAS}
    \label{fig:WaitingTimes}
\end{figure}

\begin{figure}[htb!]
    \centering
    \includegraphics{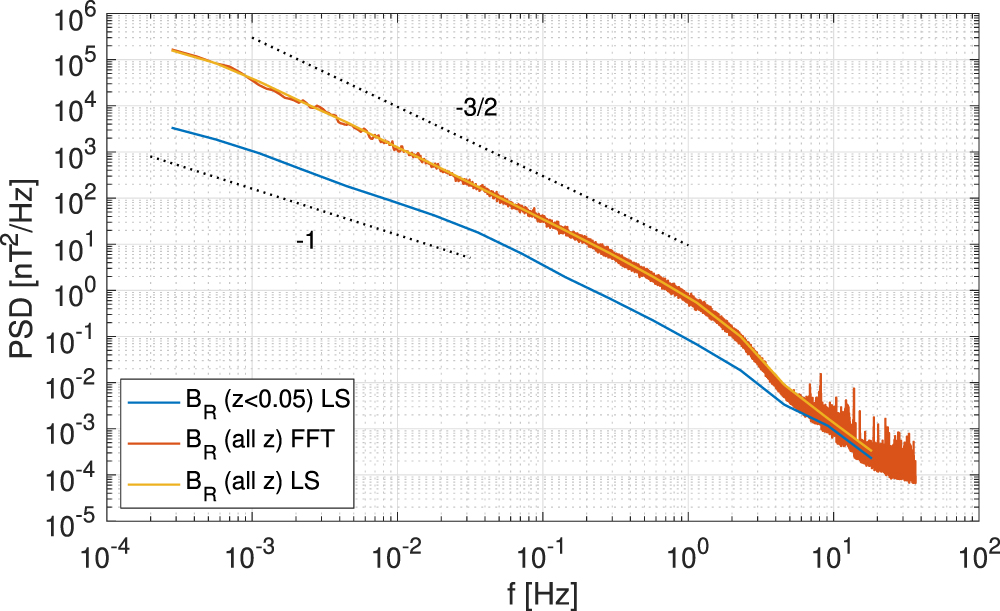}
    \caption{Power spectrum density of the radial component of magnetic field $B_R$. Blue curve corresponds to quiescent intervals, and yellow and red curves correspond to the inclusion of ``active'', i.e. switchback, intervals. Figure reproduced with permission from \citet{DudokdeWit2020}, copyright by AAS}
    \label{fig:dewit2020_fig09}
\end{figure}

In contrast, \cite{DudokdeWit2020} reported a similarity of the probability density function (PDF) of intervals between switchbacks (or ``waiting times") to that of their durations, and noted {\em ``the sharpness and omnipresence of these events, as if the coronal plasma was continuously transitioning between two metastable states \ldots "}. 
This observation may support the possible scenario that switchbacks are standalone structures.
Furthermore, it motivates revisiting the simplified models of ``intermittent switching'' explored since the 1960s by \cite{Mandelbrot1967} and others, e.g. \cite{Malakhov1993} and \cite{Niemann2013}.
Such models, reviewed by \cite{Watkins2017,Watkins2019}, have two or more states and typically have shallow power-law PDFs for the waiting times between state changes. 
In this family of models, the power spectral density is controlled solely by the PDF of waiting time rather than the PDF of amplitudes, and so it is irrelevant whether the latter is two-state or multi-state.
This robustness is desirable in view of the non-uniqueness of the definition of states, e.g. the threshold in $z$-value as used by \citet{DudokdeWit2020}.  
As shown by Figure \ref{fig:WaitingTimes}, the PDFs of waiting time for five different $z$-thresholds were found to collapse onto one single curve except for the longest intervals ($ \tau \gtrsim 2$ minutes), and they tended to follow an inverse power law $\psi \sim \tau^{-(1+\alpha)}$, with  $0.4 < \alpha < 0.6$.  
Multilevel switched models with power law waiting time PDFs  have power spectral density $S(f) \sim 1/f^{2-\alpha}$ \citep{Niemann2013}, so these observed waiting time PDFs predict a range between  $\sim f^{-1.4}$ and $\sim f^{-1.6}$, not dissimilar to the red curve (for the power spectrum of $B_R$) in Figure \ref{fig:dewit2020_fig09}. 
Because multilevel switching models typically have $\alpha$ in the range $0$ to $1$ they give power spectra from $f^{-1}$ to $f^{-2}$. They can be further adapted and generalised, e.g. to the truncated power law case studied by \cite{Lowen}.

Although whether the switchbacks are part of the turbulence or not is still under debate, many studies have shown that properties of the solar wind turbulence strongly depend on the presence of switchbacks.
Figure \ref{fig:dewit2020_fig09} shows the result using Encounter 1 data. Here, the power spectrum of the radial component of the magnetic field is evaluated using a Lomb–Scargle method \citep{macdonald1989spectral}, which is able to estimate the power spectral density from irregularly sampled data. 
For quiescent intervals (blue curve), i.e. intervals without switchbacks, the power spectrum displays a clear break frequency around 0.05 Hz that separates the spectrum into a shallower energy-containing range and a steeper inertial range. 
On the contrary, if all intervals are considered (orange and red curves), the inertial range is much more extended, with the break frequency much lower than that of the quiescent intervals. 
This indicates that turbulence is more developed in the switchbacks intervals than in the quiescent intervals because the low-frequency break point of the power spectrum is expected to move toward larger scales as the turbulence cascade proceeds. 
We note that in \citet{DudokdeWit2020} the slope of the inertial range does not change much between switchback and quiescent intervals.
\citet{Bourouaine2020} utilized the conditioned correlation functions to estimate the power spectra of different fields with encounter 1 data. They find that the magnetic field spectrum is on average steeper in the switchback intervals than the non-switchback intervals. 
The same trend was also reported by \citet{Shi2022}.
In addition to spectral slopes and the Alfv\'enicity discussed in Section \ref{sec:alfvenicity}, switchbacks may also affect the energy cascade rate of the turbulence. 
\citet{Martinovic2021} found that, while the spectral slopes and stochastic heating rates do not differ much inside and outside the switchbacks, the intermittency level is significantly stronger inside the switchbacks.
\citet{Hernandez2021} showed that broad intervals containing more switchbacks have enhanced intermittency and larger cross-helicity. The same authors calculated the turbulence energy cascade rate from the Politano–Pouquet law \citep{Marino2023} and identified a moderate correlation between the switchback amplitude and the cascade rate, implying that switchbacks contribute to injecting energy into the turbulence cascade. 

Finally, another important question is whether we expect turbulence to destroy switchbacks as they propagate out from the Sun. This has been addressed with scaling arguments by \citet{Johnston2022}, using earlier insights from \citet{dmitruk2003} and \citet{chandran2009}. Assuming that the turbulence is of the reflection-driven variety, upon balancing the source of the inward-propagating $\bm{z}^-$ (reflection) with nonlinear damping, they obtain an expression for the resulting amplitude of $z^-$. Inserting this into the equation for nonlinear evolution of $\bm{z}^+$, {\citet{Johnston2022}} obtain, for distances much above the Alfv\'en radius $R_A$,
\begin{equation}
    \frac{z^+}{v_A} \propto R^0, \quad (R\gg R_A)
\end{equation}
meaning that in this regime, turbulent damping balances the expansion-driven growth of the normalised fluctuations, and switchbacks might be expected to neither grow nor decay due to the combination of these two mechanisms. The situation is very different below the Alfv\'en radius: because expansion increases the energy in the fluctuations very fast in this regime, turbulent decay cannot overcome the growth {\citep{chandran2009}}, leading to
\begin{equation}
    \frac{z^+}{v_A} \propto \frac{U^{1/2}}{v_A},
\end{equation}
and since $U$ and $v_A$ are increasing and decreasing functions of $R$ respectively, this implies rapid growth of the fluctuation amplitude relative to the background magnetic field, possibly leading to the generation of switchbacks, even including the effects of turbulent damping.

\subsection{Solar Wind Acceleration}\label{sec:acceleration}


Under conservation of wave action, the effect of Alfv\'en waves on an expanding solar wind is to transfer net mechanical work to the plasma and leads to a larger asymptotic solar wind speed, the formalism of which was developed in the 1970s \citep[e.g. ][]{Alazraki1971,Hollweg1973,Jacques1977}. Therefore, in the absence of dissipation, a natural consequence of the presence of a substantial Alfv\'en wave energy flux at the base of the solar wind is a more steeply accelerated solar wind profile. As such, Alfv\'en waves have been previously hypothesized to play a significant role in the formation of the fast solar wind. 

With new measurements from Parker Solar Probe and Solar Orbiter, the actual acceleration profile of the solar wind down to the point where it becomes sub-Alfv\'enic is now beginning to be directly measurable \citep{Dakeyo2022,Halekas2022}. Moreover, the \emph{in situ} data collected as these profiles are measured allows for the construction of energy budgets, including the energy flux content in Alfv\'enic fluctuations \citep{Halekas2023,Rivera2024}, and for the first time therefore the relative importance of such wave energy on solar wind acceleration can be directly traced. While the impact of switchbacks on solar wind acceleration is discussed in some length in \cite{paper2}, some key recent results are worth remarking on briefly. 

\citet{Halekas2022} categorized the solar wind measurements from Parker Solar Probe into families of asymptotic wind speed using measurements of the ambipolar potential and energy conservation arguments. Using this, they derived acceleration profiles over a range of solar wind speeds and showed that for the slowest winds an exospheric or simple Parker model could explain the acceleration, i.e. no significant source of mechanical work was missing for those winds. 
Conversely, the fastest winds acceleration required a substantial missing energy flux component not present in those simplest models. 
Here, we note that the simple Parker model adopted by \citet{Halekas2022} is not fully self-consistent and assumes adiabatic index $\gamma=5/4$. In a later work by \citet{shi2022acceleration}, the authors developed a self-consistent theory for Parker-type wind with arbitrary $\gamma$. It was shown that a realistic solar wind speed cannot be achieved with $\gamma \gtrsim 5/4$, and an transonic solar wind solution does not exist for $\gamma \geq 3/2$. Thus, additional energy is necessary to generate a realistic solar wind in Parker-type solar wind models.

In a follow-up, \citet{Halekas2023} performed a similar statistical analysis and tracked the energy budget of the different families, including Alfv\'enic wave energy flux. They found that, while for the slowest solar winds again the wave energy flux was not consequential, the faster the wind the more important it was, i.e. the fastest asymptotic speeds observed by Parker Solar Probe at aphelion \textit{require} the energy input of Alfv\'en waves lower down to explain their achieved kinetic energy. 
Lastly, \citet{Rivera2024} have studied a solar wind conjunction between Parker and Solar Orbiter from February 2022 (Encounter 11) following the evolution of switchback patch from 13 to 127~$R_\odot$. Such a stream would be considered ``fast'' at 1~au but was substantially slower at Parker. This study shows that the energy flux of the switchback patch is necessary to achieve energy conservation and explain the observed acceleration profile. Further, they showed that the energy flux lost from the waves is steeper than if wave action was conserved, and that therefore switchbacks may contribute to sustaining non-adiabatic temperature profiles in the solar wind. 

Moving forward, such energy budget studies can continue deeper into the solar wind to study the acceleration properties closer to the sonic critical point. Inroads are already being made by combining Parker's \emph{in situ} data with remote observations from Solar Orbiter \citep{Telloni2023a,Telloni2023b}.

\subsection{Open Flux}

\begin{figure}
    \centering
    \includegraphics[width=0.45\textwidth]{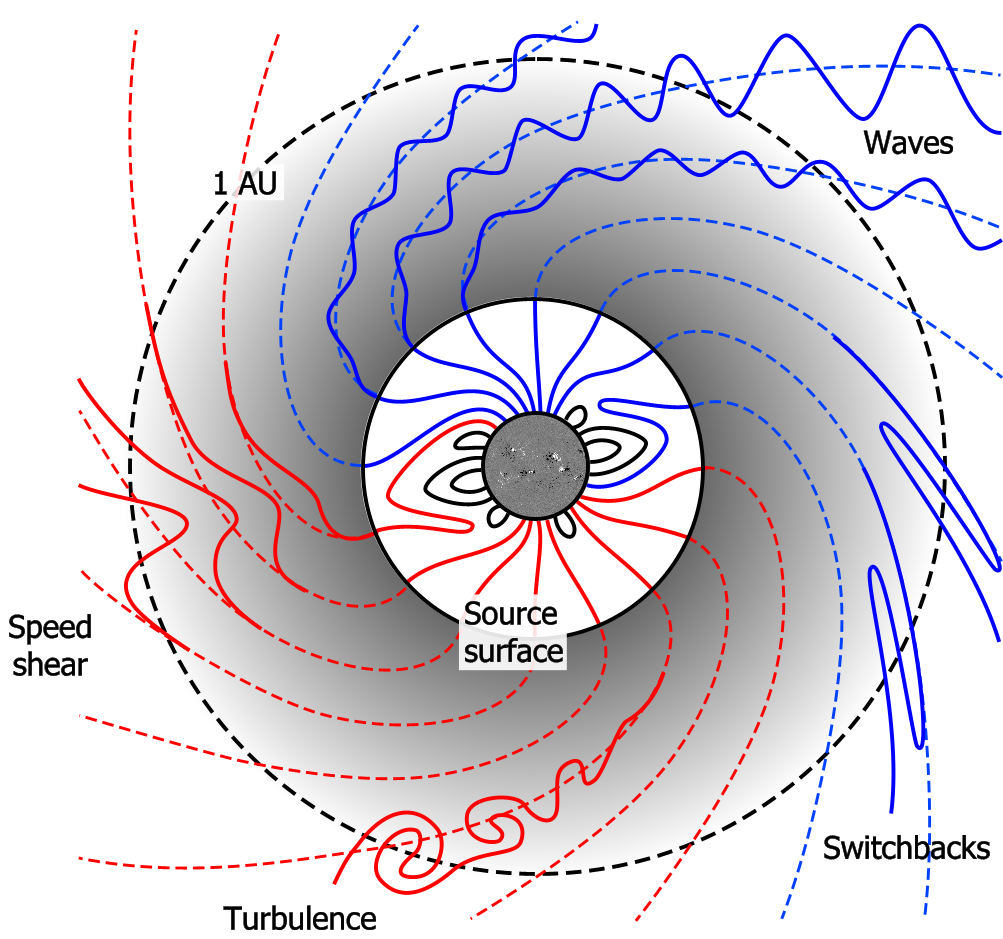}
    \caption{Schematic of the heliospheric magnetic field. The OSF is defined as the total unsigned magnetic flux threading the source surface. Estimating the OSF from heliospheric observations, such as at 1 au, is complicated by inverted HMF from waves, stream shear, turbulence and switchbacks. Figure reproduced with permission from \citet{owens2013heliospheric}, copyright by Springer}
    \label{fig:osf_cartoon}
\end{figure}

 The open solar flux (OSF) is the component of the coronal magnetic field that reaches sufficient altitude to be dragged out by the solar wind to form the heliospheric (or interplanetary) magnetic flux (HMF). 

 OSF can be estimated from photospheric magnetograms, using coronal extrapolation methods such as the potential-field source surface (PFSS) model \citep{AltschulerNewkirk1969, Schatten1969} and computing the total magnetic flux of field lines which thread the outer boundary of the model. It can also be estimated from \emph{in situ} observations of the radial component of the HMF, $B_r$, made at a heliocentric distance $R$ by computing OSF = $4 \pi r^2 |B_r|$. This assumes that local measurements of $B_r$ are representative of all positions on the sphere of radius $R$. Longitudinal variability can be removed by integrating over a solar rotation ($\approx$ 27 days, for spacecraft in near-Earth space). Latitude is accounted for by assuming no variation in $B_r$ with latitude, an assumption found to be valid to considerable accuracy by measurements from the Ulysses spacecraft \citep{smith_ulysses_1995-1, lockwood_open_2004}, and which can be understood as a vanishing (balanced) latitudinal Lorentz force established circa $10 R_S$ \citep[see e.g.,][]{suessLatitudinalDependenceRadial1996,RevilleBrun2017}.

 However, the existence of inverted HMF \citep[e.g.][]{crookerLargescaleMagneticField2004, kahlerPropertiesInterplanetaryMagnetic1998} complicates this calculation. Inverted HMF is flux which threads a sphere at the point of observation, typically $r = 1$~au, multiple times, but only threads the source surface once. For the purposes of computing OSF, this produces an ``excess flux'' at $r$ compared to the source surface. This may explain a portion of the approximately factor two difference between \emph{in situ} and photospheric OSF estimates \citep{wallaceEstimatingTotalOpen2019, linkerOpenFluxProblem2017}. 
 
\begin{figure}
\includegraphics[width=0.6\textwidth]{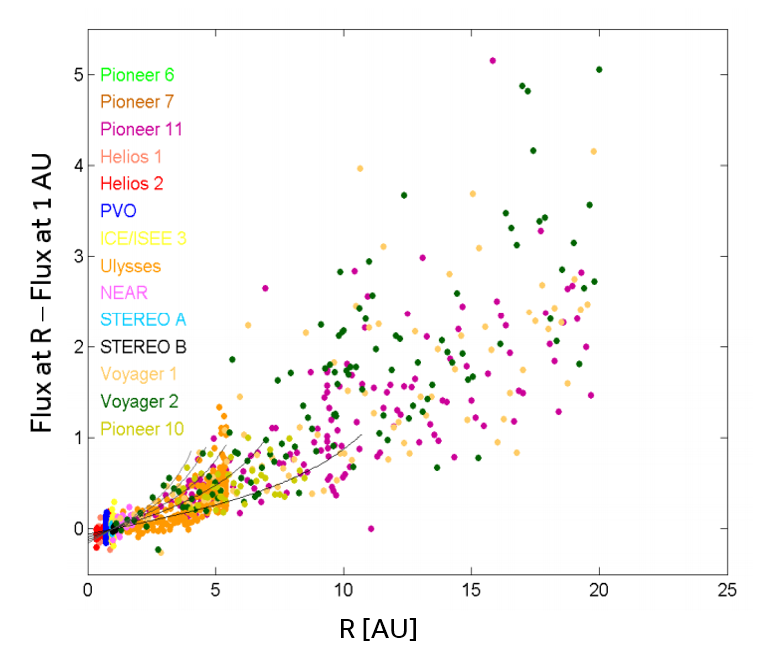}
\centering

    \caption{Difference between the total heliospheric flux (i.e. $4 \pi |B_r| R^2$) at distance $R$ and 1 au, as a function of $R$, using a range of heliospheric spacecraft over the last 60 years, using data from \cite{owensEstimatingTotalHeliospheric2008}. Reproduced with permission from \cite{lockwoodExcessOpenSolar2009}, copyright by Wiley}
    \label{fig:osf_insitu}
\end{figure}

 Excess flux will be generated by any processes resulting in ``orthogardenhose'' heliospheric flux \citep{lockwoodOriginOrthoGardenhoseHeliospheric2019}, including coronal interchange reconnection \citep{owensSolarOriginHeliospheric2013}, large-scale solar wind stream shears 
\citep{lockwoodExcessOpenSolar2009}, turbulence \citep{bruno2013solar} and, of course, kinetic-scale ``switchbacks'' \citep[e.g.][]{mozerOriginSwitchbacksObserved2021},  as shown schematically in Figure \ref{fig:osf_cartoon}. Unless inverted HMF is accounted for in some way, either explicitly or implicitly, these processes result in increasing OSF estimates with heliocentric distances, as shown in Figure \ref{fig:osf_insitu}.

It is important to note that, while a deflection greater than 90 degrees of the field is not a particularly meaningful threshold for defining switchbacks in the general context of their identification as large amplitude Alfv\'en waves (see \cite{paper3} for a discussion of definitions), in the specific application of open flux, it is an important due to the topological implication of measuring excess magnetic flux. 

For observations at 1~au, inverted HMF is generally the result of larger-scale structures, which can be identified and subtracted using, e.g., suprathermal electron observations \citep{frostEstimatingOpenSolar2022}. In the inner heliosphere, as smaller-scale Alfv\'enic switchbacks become more prevalent, this becomes more challenging -- an event-by-event based subtraction becomes intractable although the fraction of time the field is locally inverted remains a useful metric \citep{Badman2021}.

First with Helios, and now PSP allow the measurement of the filling fraction of the solar wind with inverted field lines much closer to the Sun than 1~au. To date, it has been shown  \citep{mozerOriginSwitchbacksObserved2021, macneilRadialEvolutionSunward2020, Badman2021} that the fraction of inverted flux increases steadily with heliospheric distance. Therefore, the importance of the effect will likely grow with heliospheric distance. However, \cite{Badman2021} used data over the first five orbits of Parker to estimate the total excess contribution from inverted flux as a function of distance from 1 au down to 28 $R_S$ and found it to be negligible at closest approach, while the mean measured flux \emph{in situ} remained significantly higher than coronal extrapolations, suggesting inverted flux is not the dominant reason for the open flux problem close to the Sun. This is in good agreement with the estimate of the inverted flux contribution at 1 au from a few decades of \emph{in situ} observations \citep{wallaceEstimatingTotalOpen2019, frostEstimatingOpenSolar2022}. 

\subsection{Energetic Particles}\label{sec:particles}

Irregularities in the interplanetary magnetic field ranging from small to large scales can influence particle transport; these include interplanetary magnetic turbulence \citep{MatthaeusQin2003,Pucci2016}, as well as coherent structures \citep{Tessein2015}. 
Particle transport is known to be affected by turbulence properties such as the fluctuation amplitude, the spectral index, and the anisotropy in the wave vector space \citep{Jokipii1966,MatthaeusQin2003,Pommois2005,HusseinShalchi2016}. Parallel and perpendicular transport are determined by various mechanisms, such as random walk of magnetic field lines, pitch-angle diffusion, and drift motion due to magnetic field inhomogeneities \citep{Moraal2013,Shalchi2009}.  Thus, the  marked magnetic field inhomogeneity related to switchbacks can also be expected to influence the transport of energetic particles in the heliosphere.

In considering diffusion of energetic particles in a rotational discontinuity (RD), \citet{Artemyev2020} found that the violation of the longitudinal adiabatic invariant could result in rapid pitch-angle scattering. Moreover, \citet{MalaraPerriZimbardo2021} found that particles exhibit a chaotic behaviour and can remain trapped inside the discontinuity, with trapping times displaying a nearly power-law distribution. In general, particles with gyroradii comparable to the thickness of the RD are most affected.
A similar behaviour can also be expected when energetic particles interact with switchbacks. For instance, using the EPI-Lo instrument onboard PSP, \citet{Bandyopadhyay2021} analyzed the anisotropy of $\gtrsim$100 keV protons and found that their propagation directions did not change from anti-sunward to sunward when crossing switchbacks. This allowed them to estimate upper limits for radius of curvature of the magnetic field in the observed switchbacks of $\sim$4000 km at heliocentric distances of $\gtrsim$ 28 $R_{\odot}$.

\begin{figure}
    \centering
    \includegraphics[width=0.55\textwidth]{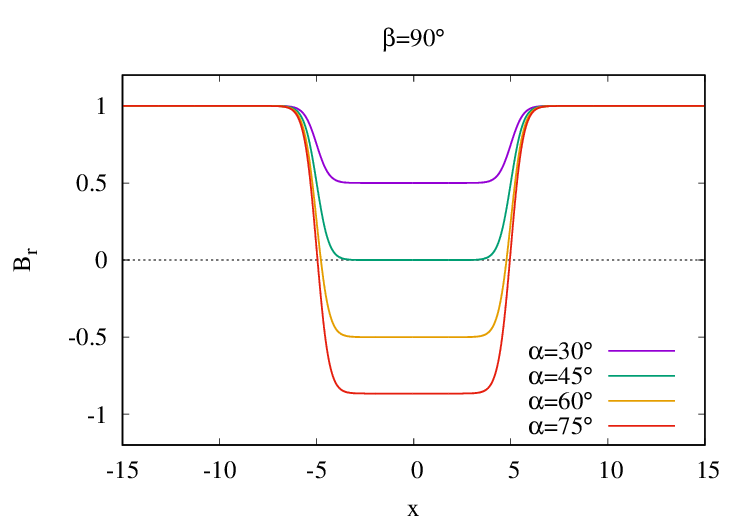}
    \caption{The main magnetic field component $B_r$ is plotted as a function of $x$, for rotation angle $\beta=90^\circ$ and various values of the inclination angle $\alpha$. The component $B_r$ reverts its sign in the central region for $\alpha > 45^\circ$. Figure reproduced with permission from \citet{Malara2023}, copyright by ESO}
    \label{Fig1:Francesco}
\end{figure}

The dynamics of energetic particles impinging on switchbacks has been studied by \citet{Malara2023}, who employed a simple 1D model for the magnetic field $\boldsymbol{B}$. In this model a static $\boldsymbol{B}$ varies only along a given spatial direction ($x$) but has a uniform magnitude in space. The switchback is limited by two opposite RDs. 
According to the values of the involved parameters (the angle $\alpha$ between $\boldsymbol{B}$ and the $x$ direction and the angle $2\beta$ giving the amount of rotation across the two RDs), the main magnetic field component (denoted as $B_r$) can display a sign inversion inside the switchback (Fig. \ref{Fig1:Francesco}). Other relevant parameters are the width $2x_c$ of the switchback itself and the width $\Delta x$ of the two RDs located at the switchback boundaries. 

A population of energetic protons in which each proton has the same energy $\mathcal{E}$ moving toward the switchback is considered. Particle time evolution is calculated by numerically solving the relativistic {equation of motion} in the background plasma reference frame. Since the energetic particle speed is typically much larger than the Alfv\'en velocity, only contributions of the magnetic force is retained. Therefore, $\mathcal{E}$ remains constant in time. The direction of the initial particle velocity is uniformly distributed on a hemisphere centered around the $\boldsymbol{B}$ direction far from the switchback. The proton energy $\mathcal{E}$ was varied in a range between 100 keV and 3 GeV. Values assumed for other parameters were \citep[e.g.,][]{Pecora2022}: magnetic field intensity $B=15$ nT; RD crossing time $\delta t=28$ s (in the spacecraft reference frame); solar wind velocity $v_{SW}=340$ km s$^{-1}$, giving a RD width $\Delta x\sim 9500$ km {consistent with estimated values of boundary thickness \citep[e.g.,][] {Larosa2021, Bizien2023}}; switchback half width $x_c$ varying between $x_c=5 \Delta x$ and $x_c=15 \Delta x$. Using the above value of $B$ the proton Larmor radius $\rho$ varies between $\rho\simeq 0.3 \Delta x$ for $\mathcal{E}=100$ keV, up to $\rho \gg x_c$ for $\mathcal{E}$ of the order of a few GeV. However, the range of variation of these parameters can be broad \citep{DudokdeWit2020,Pecora2022,Larosa2021, Bizien2023}.

\begin{figure}
    \centering
    \includegraphics[width=0.55\textwidth]{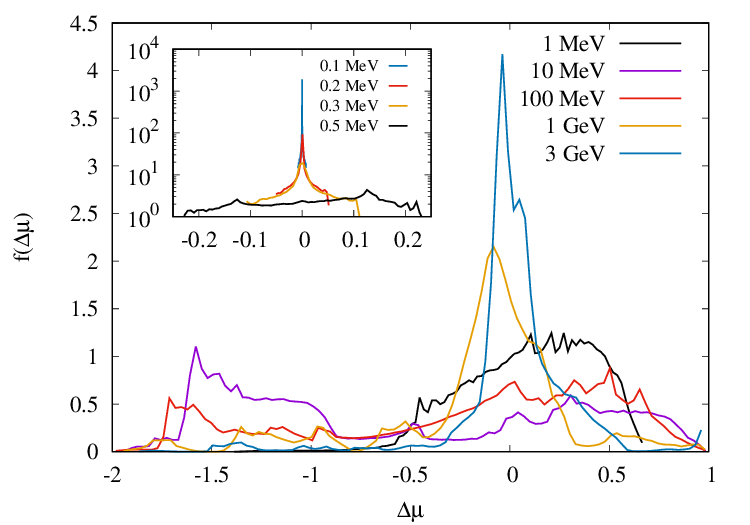}
    \includegraphics[width=0.55\textwidth]{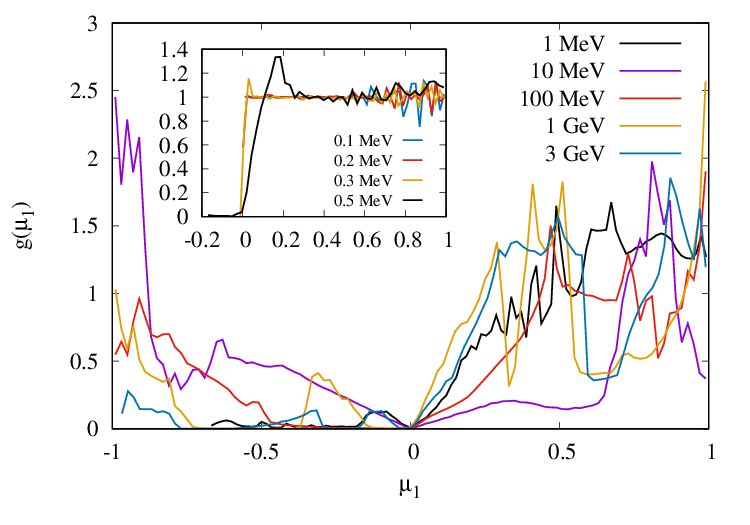}
    \caption{Distributions $f(\Delta \mu)$ of pitch-angle cosine variations (top panel) and $g(\mu_1)$ of final pitch-angle cosine (bottom panel) calculated for various values of the particle energy $\mathcal{E}$ in the range $0.1$ MeV $\le \mathcal{E} \le 3 $ GeV. The cases $\mathcal{E} = 0.1, 0.2, 0.3, 0.5$ MeV are shown in the insets. Both panels refer to a magnetic structure where $\beta = 90^\circ$ and $\alpha = 60^\circ$. Figure reproduced with permission from \citet{Malara2023}, copyright by ESO}
    \label{Fig2:Francesco}
\end{figure}

The magnetic field inhomogeneity associated with the switchback structure causes pitch-angle scattering in the particle population. The intensity and properties of such a phenomenon depend both on the parameters defining the switchback structure and on the particle energy that determines the Larmor radius. Pitch-angle scattering is illustrated in Fig. \ref{Fig2:Francesco} (upper panel), where the distribution $f(\Delta \mu)$ of pitch-angle-cosine variations $\Delta \mu=\mu_1-\mu_0$ is plotted, $\mu_0$ and $\mu_1$ being the cosines of initial and final particle pitch angle, respectively. The quantity $\mu_1$ is evaluated when the particle has completely left the switchback. For low energies, of the order of a few keV, the proton Larmor radius is smaller than the RD width. Therefore, the particle magnetic momentum is conserved. Along with energy conservation, this condition leads to constant pitch-angles, as shown in Fig. \ref{Fig2:Francesco} (inset in upper panel). Lower level of pitch-angle scattering is also present at higher energies; e.g., for $\mathcal{E} \gtrsim 1$ GeV the distribution $f(\Delta \mu)$ has a smaller spread or full-width half maximum, centered around $\Delta \mu=0$. For those energies, the particle Larmor radius is much larger than the switchback size; therefore, protons ``leap over'' the switchback, thus minimizing their interactions with the magnetic field inhomogeneity. As a consequence, their pitch angles are nearly conserved. 
For intermediate energies between $1$ MeV $\lesssim \mathcal{E} \lesssim 100$ MeV, the proton Larmor radius is of the order of the RD width $\Delta x$ or of the SB size. In this case,  effects of the $\boldsymbol{B}$ inhomogeneity are important, as the interactions cause significant pitch-angle scattering, which is characterized by broader distributions of $f(\Delta \mu)$.

Distributions $g(\mu_1)$ of the final pitch-angle cosine $\mu_1$ are plotted in the lower panel of Fig. \ref{Fig2:Francesco} for various values of $\mathcal{E}$. In the intermediate energy range, the formation of a population of back-reflected particles is represented by negative values of $\mu_1$. In this case, a certain fraction of particles interacting with the switchback returns in the opposite direction. A parameter that gives a measure of the effect of the switchback on the proton dynamics is the ratio $R=N_{refl}/N_{tot}$ giving the fraction of reflected particles. Depending on the choice of model parameters, $R$ could get as high as $60-70 \%$ for protons with energies $\mathcal{E}\simeq 10-100$ MeV. 
Another feature of the distribution $g(\mu_1)$ is the prevalence of values of $\mu_1$ close to $\mu_1 = \pm 1$. This indicates a tendency of particles to converge their velocities in directions that are nearly parallel or anti-parallel to the magnetic field. Concerning particle dynamics, depending on their initial pitch angle and gyrophase, some particles remain temporarily trapped inside the switchback, bouncing back and forth between the two RDs. {This complex} behaviour is extremely sensitive to initial conditions. 

Results of the above model indicate some possible consequences from the observational point of view {that maybe investigated in future observational studies}. For instance, if a solar energetic particle (SEP) event is impinging on an SB that is located farther away from the Sun, then a certain fraction of these particles can be scattered back towards the Sun. This would give rise to a decrease of the intensity of SEPs beyond the switchback. The energy range of such dropouts in the energetic particle fluxes can be predicted if the SB size is obtained from the measured SB duration and the solar wind speed. Moreover, transmitted particles may become more field aligned, as shown by the distribution of final pitch-angle cosines. These properties could be checked by multi-spacecraft observations, if simultaneous SEP measurements by magnetically connected spacecraft on both sides of a switchback are available. Finally, the strong pitch-angle scattering in particles crossing the SB can have an influence on the processes of Fermi acceleration, both first and second order. In fact, switchbacks can act as magnetic mirrors, even if the magnetic field magnitude is constant. These effects should be taken into account when studying energetic particle propagation and acceleration in the heliosphere.

\subsection{Heating of the Solar Wind by Switchbacks}\label{sec:heating}

Understanding the heating of the solar wind and the corona remains a primary objective of heliospheric physics. 
Statistical surveys in the inner heliosphere confirm that the plasma heating associated with turbulent fluctuations increases significantly closer to the Sun \citep{Wu2022,Sorriso-Valvo2023}. 
It is still unclear if and how the enhanced presence of switchbacks contributes to such heating. It is, however, understood that the energy and momentum stored in the switchbacks must be conserved and dissipated into the plasma, likely generating heating and particle energization. 
Understanding the evolution of large-amplitude Alfvénic switchbacks and the processes by which they dissipate energy is therefore necessary in determining their role in solar wind heating, and may provide constraints on the general heating in the solar wind. 

As previously discussed, using the Politano-Pouquet third-order moment scaling law, \citet{Hernandez2021} and \citet{Marino2023} measured the turbulence energy transfer rate in intervals with various levels of occurrence of switchbacks. 
Although the study was performed on a limited ensemble (the first PSP encounter), it was found that the occurrence of switchbacks positively correlated with the energy transfer rate.
This indicates that switchbacks are associated with enhanced transfer of turbulent energy at small scales, where it is available to be converted into heating via kinetic processes.

Observations from Helios suggest that proton populations undergo perpendicular heating in the inner heliosphere, where switchbacks are most common \citep{Marsch1982b,Hellinger2011}.  
Recent work from PSP has largely reproduced these results, although the inclusion of empirical measurements rather than power-law scaling functions has suggested that the parallel temperature may follow adiabatic scaling \citep{Zaslavsky2023,Mozer2023a}. In either case, it is well established that a significant fraction of heating should go into the perpendicular ion population.

The partition of heating between protons, electrons, and heavy ions is a primary signature that can help understand how heating happens. 
\citet{Bandyopadhyay2023} apply the \citet{Cranmer2009} turbulence model to estimate heating rates for the ions and electrons, demonstrating that at closer heliospheric distances ion heating is dominant. 
It is at these regions where switchbacks are prominent, indicating that {either} there {is} a preference for ion-scale heating from the Alfvénic switchbacks, {or that the switchbacks and the ion heating have a common cause.}

\citet{Shankarappa2023} obtained similar heating rates within the \citet{Howes2008} cascade model, which relies on Landau damping, and is sensitive to plasma $\beta$. 
Slight differences in the heating rates observed by \citet{Bandyopadhyay2023}, which is agnostic to the heating mechanism, and \citet{Shankarappa2023}, which only accounts for Landau damping, suggest that alternative mechanisms may play a dominant role in proton heating in the inner heliosphere.

The application of quasilinear theory \citep{KennelEngelmann1966} to observed populations of cyclotron waves 
has shown cyclotron resonant heating to be active in intervals with switchbacks \citep{Bowen2022}, accounting for 10-25\% of the turbulent cascade rate. Furthermore \citet{Bowen2024} show that Alfv\'enic fluctuations with large cross-helicity are preferentially associated with the generation of ion cyclotron waves. 
Signatures of ion and sub-ion scale turbulence, including spectral slopes and intermittency, show that  kinetic scale turbulence is strongly mediated by the presence of waves, indicating that dissipation of strongly Alfv\'enic fluctuations, such as the switchbacks, may be dissipated at ion scales via cyclotron resonance. 
In less Alfv\'enic intervals, the turbulence seems to cascade to sub-ion scales where it may be dissipated via current sheets. 

Finally, a possible association between switchbacks and solar wind heating seems to be corroborated by a recent result obtained using Wind spacecraft observations of a CME sheath at 1~au.
\citet{Yordanova2021} found good correlations between plasma heating and enhanced small-scale current and vorticity Alfv\'enic structures.
The study revealed that such turbulent heating manifests in clusters, or ``blobs'', whose typical scale seems to compare well with the size of patches of switchbacks observed near the Sun (see Section \ref{sec:observation_occurrence}). 
The possible role of switchback patches in modulating the mesoscale solar-wind turbulent heating is currently under study.

\section{Summary and Future Directions}

Since Parker Solar Probe provided evidence that switchbacks are an intrinsic characteristic of the young solar wind  \citep{Bale2019,Kasper2019}, much of the early work has been focused on characterizing their properties \citep[see][]{paper3} and how they might be generated \citep[see][]{paper5}.
In contrast, while many promising studies and lines of inquiry have been started (as can be seen in this paper), a comprehensive physical understanding of how switchbacks evolve and their impact on the solar wind as a whole has yet to be reached.

In this paper, we have summarized the emerging lines of inquiry that have been started to explore this topic.
Simulations and observations show that the solar wind expansion may indeed favor the generation of switchbacks from smaller-amplitude seed fluctuations \emph{in situ}, but their evolution appears to be complex and scale-dependent.
Depending on their topology and the background solar wind parameters, switchbacks may undergo steepening, dispersion, reconnection, as well as parametric decay instability. 
As they evolve in interplanetary space and disrupt or dissipate, switchbacks potentially contribute to the internal energy of solar wind, though how they impact the turbulent cascade remains an open issue due to their highly Alfv\'enic nature.

Despite the established understandings mentioned above, many questions remain unanswered.
This is in part due to the difficulty of studying the evolution of structures that are generally only observed once -- the usual problem with single-spacecraft measurements of solar wind phenomena. 
Moreover, realistic numerical simulations of switchback evolution are necessarily limited due to the enormous scale separation between the inhomogeneities at the system scale (a few tens of solar radii) and the small scales relevant to the switchbacks, particularly their boundaries, which can be on the order of hundreds of kilometers.
Indeed, so far, there has been no simulation that has successfully generated a sufficient number of switchbacks self-consistently \citep{Squire2020,Shoda2021}, possibly due to the lack of spatial resolution.

However, although it is impossible to trace the evolution of one switchback with a single spacecraft, with the enormous number of switchbacks observed by PSP, we are able to conduct statistical investigation, which has been adopted in most of the previous studies \citep[e.g.][]{Horbury2020,DudokdeWit2020,Tenerani2021}.
Additionally, although global simulations of the switchback evolution are difficult, high-resolution local simulations can be remarkably valuable in investigating this problem.
Specifically, the expanding-box-model \citep[EBM,][]{grappin1996waves} enables us to trace the evolution of a small parcel of expanding solar wind and has been implemented in recent MHD simulations \citep{Squire2020,Johnston2022} and hybrid simulations \citep{matteini2024alfvenic} to study the generation and evolution of switchbacks. 
Therefore, numerous potentially important studies should be conducted in the near future, several of which are listed below:
\begin{itemize}
    \item Understanding the 3D topology/geometry of the switchbacks. Previous work has shown that switchbacks are stretched structures along the background magnetic field \citep{Horbury2020,Laker2021}, and numerical simulations support this observation as the elongated switchbacks are more stable than shorter ones \citep{shi2024analytic}. 
    Nonetheless, we still do not have a clear picture of the switchbacks' topology in 3D, and how their topology depends on radial distance to the Sun and other solar wind parameters. A comprehensive statistical analysis of all the available PSP measurements will be necessary.
    \item Understanding whether there are different types of switchbacks. The widely accepted definition of a magnetic switchback, i.e. a local polarity reversal of the magnetic field, is broad. 
    However, in practice, switchbacks can be categorized based on parameters such as their compressibility, Alfv\'enicity, generation and dissipation mechanisms, etc.
    Different types of switchbacks may undergo significantly different evolution. 
    Thus, it will be necessary to categorize the switchbacks instead of adopting a single, universal definition, and to investigate the evolution of different types of switchbacks.
    \item {Quantifying the contribution of switchbacks to the heating and acceleration of solar wind. Recent energy budget statistics \citep{AkhavanTafti2022,Halekas2023} and spacecraft conjunction studies \citep{Rivera2024,Soni2024arXiv240213964L,rivera2025differentiating} have provided strong evidence that switchbacks, or Alfv\'enic turbulence in general, are a significant energy flux term in fast solar wind close to the Sun, which must ultimately be transferred to the bulk flow of the plasma. However, the physical mechanism for how this energy is transferred and to what extent it is partitioned into heat or kinetic energy (or mediated by other intermediate states such as compressive fluctuations) remains to be established. A promising method to address this is by analyzing the ``fast radial scan'' intervals of PSP when it travels almost in a purely radial direction and measures stream evolution directly.}
    \item Understanding the early-stage evolution of switchbacks. 
    It is likely that the switchbacks are generated \emph{in situ} due to the solar wind expansion \citep{Squire2020,Shoda2021} while the ``seed'' fluctuations may be injected in the lower solar atmosphere through a wide range of different processes \citep{norbert2021,Finley22,Drake2021,Wyper2022,touresse2024}. 
    How these seed fluctuations eventually develop into switchbacks, and what physical processes control this evolution are important questions but are not fully understood yet. For example, why and how do the fluctuations evolve into a uniform-$|B|$ status \citep{matteini2024alfvenic}?
    Comprehensive numerical investigations will be necessary to answer these questions. For further details of the current state of the art, \citet{paper4} reviews the solar physics processes that could contribute to switchbacks, while \citet{paper5} reviews many proposed \emph{in situ} and lower solar atmosphere switchback generation processes. 
\end{itemize}

\backmatter





\bmhead{Acknowledgments}

The authors thank the International Space Science Institute (ISSI) for hosting the workshop on ``Magnetic Switchbacks in the Young Solar Wind'' (18-22 September 2023).
MAT was supported by NASA contract Nos. NNN06AA01C, 80NSSC20K1847, 80NSSC20K1014, and 80NSSC21K1662. AL is supported by STFC Consolidated Grant ST/T00018X/1. CS is supported by NASA ECIP \#80NSSC23K1064.  MM acknowledges DFG grants WI 3211/8-1 and WI 3211/8-2, project number 452856778. MM was also supported by the Brain Pool program funded by the Ministry of Science and ICT through the National Research Foundation of Korea (RS-2024-00408396). AM is supported by NASA grants 80NSSC21K0462 and 80NSSC21K1766. AT acknowledges support by NSF CAREER award 2141564. 
LSV is supported by the Swedish Research Council (VR) Research Grant N. 2022-03352. 
LM, MV, and LSV are supported by the International Space Science Institute (ISSI) in Bern, through the ISSI International Team project \#23-591 (Evolution of Turbulence in the Expanding Solar Wind).
OVA is supported by NASA grants 80NSSC21K1770, 80NSSC20K0218, and 80NSSC22K0433. The authors thank the International Space Science Institute (ISSI) for hosting the workshop on ``Magnetic Switchbacks in the Young Solar Wind'' (18-22 September 2023).

\bmhead{Statements and Declarations}

\bmhead{Competing Interests} The authors have no competing interests to declare that are relevant to the content of this article.

\bibliography{sn-bibliography}

\end{document}